\documentclass[twocolumn,twocolappendix]{aastex631}

\usepackage{multirow}
\usepackage[version=4]{mhchem}
\usepackage{booktabs}
\usepackage[capitalise]{cleveref}
\makeatletter
\usepackage{etoolbox}
\patchcmd\H@refstepcounter{\protected@edef}{\protected@xdef}{}{}
\makeatother
\received{20th June 2023} 
\accepted{3rd September 2024} 

\submitjournal{ApJ}

\shorttitle{SPT2349$-$56: Kinematic analysis}
\shortauthors{Venkateshwaran et al.}

\graphicspath{{./}{figures/}}

\begin{document}

\title{Kinematic analysis of $\mathbf{z = 4.3}$ galaxies in the SPT2349$-$56 protocluster core}

\author[0009-0000-9606-4380]{Aparna Venkateshwaran}
\correspondingauthor{Aparna Venkateshwaran}
\affiliation{Max-Planck-Institut für Radioastronomie, Auf dem Hugel 69, Bonn, Germany}
\email{aparnavenkat93@gmail.com}

\author[0000-0003-4678-3939]{Axel Weiss}
\affiliation{Max-Planck-Institut für Radioastronomie, Auf dem Hugel 69, Bonn, Germany}

\author[0000-0002-3187-1648]{Nikolaus Sulzenauer}
\affiliation{Max-Planck-Institut für Radioastronomie, Auf dem Hugel 69, Bonn, Germany}

\author[0000-0001-6459-0669]{Karl Menten}
\affiliation{Max-Planck-Institut für Radioastronomie, Auf dem Hugel 69, Bonn, Germany}

\author[0000-0002-6290-3198]{Manuel Aravena}
\affiliation{Instituto de Estudios Astrofísicos, Facultad de Ingeniería y Ciencias, Universidad Diego Portales, Av. Ejército 441, Santiago, Chile}

\author{Scott C. Chapman}
\affiliation{Department of Physics and Atmospheric Science, Dalhousie University, Halifax, NS, B3H 4R2, Canada}
\affiliation{NRC Herzberg Astronomy and Astrophysics, 5071 West Saanich Rd, Victoria, BC, V9E 2E7, Canada}
\affiliation{Department of Physics and Astronomy, University of British Columbia, Vancouver, BC, V6T1Z1, Canada}
\affiliation{Eureka Scientific Inc, Oakland, CA 94602, USA}

\author[0000-0002-0933-8601]{Anthony Gonzalez}
\affiliation{Department of Astronomy, University of Florida, 211 Bryant Space Science Center, Gainesville, FL 32611-2055, USA}

\author[0000-0002-7472-7697]{Gayathri Gururajan}
\affiliation{Scuola Internazionale Superiore Studi Avanzati (SISSA), Physics Area, Via Bonomea 265, 34136 Trieste, Italy}
\affiliation{IFPU-Institute for Fundamental Physics of the Universe, Via Beirut 2, 34014 Trieste, Italy}

\author[0000-0003-4073-3236]{Christopher C. Hayward}
\affiliation{Center for Computational Astrophysics, Flatiron Institute, 162 Fifth Avenue, New York, NY 10010, USA}

\author[0009-0008-8718-0644]{Ryley Hill}
\affiliation{Department of Physics and Astronomy, University of British Columbia, Vancouver, BC, V6T1Z1, Canada}

\author[0000-0001-7477-1586]{Cassie Reuter}
\affiliation{Department of Astronomy, University of Illinois, 1002 West Green St., Urbana, IL 61801, USA}

\author[0000-0003-3256-5615]{Justin S. Spilker}
\affiliation{Department of Physics and Astronomy and George P. and Cynthia Woods Mitchell Institute for Fundamental Physics and Astronomy, Texas A\&M University, 4242 TAMU, College Station, TX 77843-4242, USA}

\author[0000-0001-7192-3871]{Joaquin D. Vieira}
\affiliation{Department of Astronomy, University of Illinois, 1002 West Green St., Urbana, IL 61801, USA}
\affiliation{Center for AstroPhysical Surveys, National Center for Supercomputing Applications, 1205 West Clark Street, Urbana, IL 61801, USA}



\begin{abstract}

SPT2349$-$56 is a protocluster discovered in the 2500 deg$^2$ South Pole Telescope (SPT) survey. In this paper, we study the kinematics of the galaxies found in the core of SPT2349$-$56 using high-resolution (1.55 kpc spatial resolution at $z = 4.303$) redshifted [C\textsc{ii}] 158-$\micron$ data. Using the publicly available code \texttt{\textsuperscript{3D}BAROLO}, we analyze the seven far-infrared (FIR) brightest galaxies within the protocluster core. Based on conventional definitions for the detection of rotating discs, we classify six sources to be rotating discs in an actively star-forming protocluster environment, with weighted mean $V_{\mathrm{rot}}/\sigma_{\mathrm{disp}} = 4.5 \pm 1.3$. The weighted mean rotation velocity ($V_{\mathrm{rot}}$) and velocity dispersion ($\sigma_{\mathrm{disp}}$) for the sample are $ 357.1 \pm 114.7$ km s$^{-1}$ and $43.5 \pm 23.5$ km s$^{-1}$, respectively. We also assess the disc stability of the galaxies and find a mean Toomre parameter of $Q_\mathrm{T} = 0.9 \pm 0.3$. The galaxies show a mild positive correlation between disc stability and dynamical support. Using the position-velocity maps, we find that five sources further classify as disturbed discs, and one classifies as a strictly rotating disc. Our sample joins several observations at similar redshift with high $V/\sigma$ values, with the exception that they are morphologically disturbed, kinematically rotating and interacting galaxies in an extreme protocluster environment.

\end{abstract}





\section{Introduction} \label{sec:introduction}

Galaxy formation is thought to be driven either by `hot' or `cold' processes. In the former scenario, in-falling gas is shock-heated to virial temperatures, accreted spherically by the disc, and cooled over long timescales, eventually triggering star-formation that is regulated by feedback from stars or active galactic nuclei (AGN; \citealp{Rees_1977, White_1978, Fall_1980, King_2015}). Theoretically, this implies an increase in turbulence within galaxies at earlier times since systems are dominated by violent disc instabilities (in other words, they are dynamically hotter at earlier epochs; \citealp{Dekel_2014, Zolotov_2015, Hayward_2017, Krumholz_2018, Pillepich_2019}). 

`Cold' processes, however, describe efficient (and possibly co-planar) gas accretion via mergers or filamentary structures in the cosmic web \citep{Keres_2005, Dekel_2006, Dekel_2009a, Dekel_2009b, Kretschmer_2022}. With infrequent mergers over long time scales, this process allows relatively cold ($V/\sigma \ga 4$) discs to be present across all epochs \citep{Lelli_2016, Dekel_2020a}. Several observations have demonstrated the existence of rotational discs with $V/\sigma > 3$ at $z>2$ \citep{Hodge_2012, Lelli_2018, Sharda_2019, Rizzo_2020, Lelli_2021, Rizzo_2021, Fratenalli_2021}, which most numerical simulations have struggled to reproduce \citep{Grand_2017, Pillepich_2019}. More recently, a study using high-resolution zoom-in cosmological simulations reproduced the presence of non-transient, massive ($M_{\star} > 10^{10}$ M$_\sun$), super-cold ($V/\sigma > 10$) discs at $z \ge 4$ \citep{Pallottini_2022, Kohandel_2023}. Understanding the kinematics of such galaxies across cosmic epochs allows for studying the physical processes involved in their mass accretion. 

Studying the kinematics of galaxies at high redshift requires high-resolution data. At $z < 3$, the use of optical recombination and forbidden lines such as H$\alpha$, [OIII], and [NII] is common \citep{Swinbank_2009, Foerster_Schreiber_2009, Wisnioski_2015, Teodoro_2016, Turner_2017}. The rotational transitions of the carbon monoxide (CO) molecule are also useful for studying disc kinematics in the innermost parts of a galaxy. However, at $z \ga 3$, these lines are difficult to detect and require long integration times to reach the required sensitivities \citep{Hodge_2012, Ginolfi_2020}. Improvements in adaptive optics (AO) techniques \citep{Foerster_Schreiber_2018} and in interferometric facilities  (ALMA, ATCA, NOEMA, VLA, etc.) have significantly contributed towards improving the sensitivity and angular resolution required in such studies.

A commonly studied emission line in astronomy is that of singly ionized carbon ($\mathrm{C}^+$, hereafter [C\textsc{ii}]). Occurring due to a transition between the fine-structure levels \ce{^2P_3/2} $\rightarrow$ \ce{^2P_1/2}, it is one of the dominant cooling lines in the cool interstellar gas \citep{De_looze_2011, Wagg_2012, Herrera_Camus_2015, Lagache_2018}. [C\textsc{ii}] emission traces the atomic and molecular gas across the different phases of the interstellar medium (ISM) due to its ionization potential being lower than that of neutral hydrogen \citep{Carilli_2013}. For solar metallicity gas, it is also the brightest emission line from far-IR (FIR) to radio wavelengths. At $z \ga 2$, it shifts into the (sub-)millimeter wavelength regime, becoming a popular candidate for studying the ISM at high-redshift \citep{Posses_2023}. Due to its brightness, the line is also used in obtaining high-resolution data such as those presented in \citet{Bethermin_2020} and \citet{Walter_2022}.

Galaxy cluster progenitors, also known as `protoclusters' are found in the $z>1$ epoch overlapping with the peak in the cosmic star-formation rate (SFR; \citealp{Madau_2014, Overzier_2016}). \citet{Miley_2008} state that these protoclusters host massive starburst galaxies and are characterized by their large far-infrared (FIR) luminosities \citep{Casey_2014}. The way protoclusters are found differs from their virialized counterparts, as they lack an established intracluster medium and have a much larger angular extent. Protoclusters are usually found via examining the environments of tracer populations such as quasi-stellar objects (QSOs, i.e., radio galaxies and quasars), Ly$\alpha$ emitters, and dusty star-forming galaxies, also known as sub-millimeter galaxies (SMGs; \citealp{Weiss_2009, Dannenbauer_2014, Overzier_2016}). In recent years, surveys using the South Pole Telescope (SPT; \citealt{Carlstrom_2011}) and the \textit{Herschel Space Observatory} \citep{Herschel_2010} have revealed several rare, strongly lensed and a few un-lensed millimeter selected galaxies \citep{Eales_2010, Vieira_2010, Williamson_2011, Mocanu_2013, Oteo_2018}. Among these, SPT2349$-$56 is the brightest un-lensed protocluster core discovered at $z = 4.303$. 

Since its discovery, SPT2349$-$56 has been the subject of several follow-up observations. \citet{Miller_2018} presented ALMA (Atacama Large (sub-)Millimeter Array; \citealt{Wootten_2009}) Cycle 3 and 4 data targeting the [C\textsc{ii}] and CO(4-3) transitions, where they identified 14 galaxies at the same redshift within a radius of 65 kpc, known as the `core' region. They report an estimated SFR of around 6000 $\mathrm{M_{\sun}\,yr^{-1}}$, making the system the most active protocluster core discovered to date. Further estimates provided in \citet{Hill_2020} show the presence of over 30 galaxies contained within a 500 kpc region, with SFR of over 10,000 $\mathrm{M_{\sun}\,yr^{-1}}$. The protocluster is expected to be a Coma-type progenitor, with a cluster mass of $M > 10^{15}\,\mathrm{M_{\sun}}$ at $z = 0$ \citep{Chiang_2013, Miller_2018}.

\citet{Rotermund_2021} presented \textit{Gemini}-S (optical) and \textit{Spitzer}-IRAC (near-infrared) observations of SPT2349$-$56, identifying nine counterparts to existing galaxies and four additional Lyman-break galaxies (LBGs). The protocluster core has an estimated stellar mass of $10^{12}$ M$_\sun$, comparable to a $z = 1$ brightest cluster galaxy (BCG). With minimal detection in the optical wavelengths, the protocluster system appears to be predominantly composed of heavily dust-obscured galaxies. More recently, \citet{Hill_2022} presented deeper optical and infrared observations using the \textit{Hubble Space Telescope} and \textit{Spitzer}-IRAC, showing that the galaxies in the core are broadly aligned with the galaxy main sequence, resembling the distribution of field galaxies at a similar redshift. Non-cosmological hydro-dynamical simulations predict that the core region will collapse into a system resembling a low-redshift, giant elliptical galaxy within 100--300 Myr, depending on the true gas fraction \citep{Rennehan_2020}. 

The analysis of high-redshift galaxy kinematics is a vital component in examining galaxy evolution. The studies presented thus far have no known associations with their environments, i.e. they are `isolated' or `field' galaxies. This leaves a gap in our understanding of how galaxies behave in a turbulent and interactive environment like that of a protocluster core. In contrast to previous studies, in this paper, we present the kinematics of galaxies within the core region of the protocluster SPT2349$-$56.

\begin{figure*}
    \epsscale{1}
    \plotone{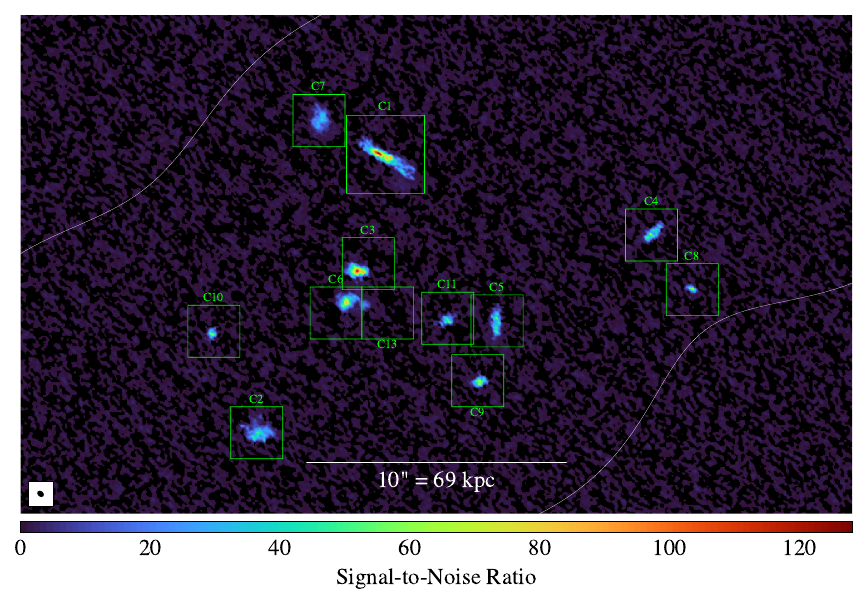}
    \caption{Signal-to-noise ratio (SNR) presentation of the continuum subtracted and primary-beam corrected integrated [C\textsc{ii}] intensity (moment-0) distribution of the core region in SPT2349$-$56. The sources are named according to \citet{Hill_2020}, and the green boxes indicate the cut-out boundaries for each source. The white contour indicates where the primary beam response falls to 80\%. The black areas are at S/N $<0$. See \cref{sec:observations_and_data_reduction} for a description of how the map was prepared.}
    \label{fig:spt2349_mosaic}
\end{figure*}

In this paper, \cref{sec:observations_and_data_reduction} describes the observations and data reduction process, \cref{sec:methods} describes the kinematic fitting process, and \cref{sec:results} presents the results from the kinematic analysis. \cref{sec:discussion} discusses the results, comparing it to similar studies across redshifts, and explores a new classification method using the position-velocity maps. Lastly, \cref{sec:conc} summarizes our findings. Throughout the analysis, we use a $\Lambda$CDM cosmology with $h = 0.678$, $\Omega_\mathrm{m} = 0.308$, $\Omega_\Lambda = 0.692$, and $1\arcsec = 6.897$ kpc \citep{Planck_2016}. 

\section{Observations and data reduction} \label{sec:observations_and_data_reduction}

We use high-resolution ALMA Cycle 6 (2018.1.00058.S; PI: S. Chapman) Band 7 data, centered on the [C\textsc{ii}] 158-$\micron$ line. The data were obtained as a 6-point mosaic over several days in November 2018, with a total on-source integration time of 35 to 57 minutes for each observation. The array configurations had baselines between 15 m and 1.4 km. The quasar J2357$-$5311 was used as a phase calibrator, while J2258$-$2758 and J2056$-$4714 were used as band-pass calibrators. We note here that this data set exclusively targets the core region presented in \citet{Hill_2020}. See Table 1 in \citet{Hill_2020} for more details about the observations.

The data were reduced using Common Astronomy Software Applications (\texttt{CASA}; \citealt{McMullin_2007}). They were concatenated and imaged using the \texttt{tclean} task, first using the multi-frequency synthesis (MFS) routine to obtain a continuum image using channels without [C\textsc{ii}] emission. We used the Briggs weighting mode and compared the results from a range of robust values between -2 and 2. We found that a robust value of 0.5 optimally balanced resolution and sensitivity. Applying the same weighting parameters, we construct a spectral cube, and subtract the continuum using the \texttt{imcontsub} task. 

The line cube was cleaned to a threshold of 0.55 mJy (1.4$\times$ the root-mean-square (RMS) noise in the dirty cube), but we note that choice of this parameter does not have a significant impact on the resulting morphology of the [C\textsc{ii}] maps. \cref{fig:spt2349_mosaic} shows the signal-to-noise ratio (SNR) map of the integrated [C\textsc{ii}] (moment-0) intensity distribution obtained using the continuum-subtracted spectral cube, averaged between $-1250$ km s$^{-1}$ and $1350$ km s$^{-1}$. 

The noise at each position was estimated using the emission free channels. A SNR cube was computed by dividing each channel map of the cube by this noise map. This approach is justified since the noise is mainly a function of position and almost constant along the frequency axis (as verified at the emission free positions). The SNR moment-0 was computed by summing all consecutive channels with SNR $> 2.5$ for positions with significant emission, while taking the noise reduction due to the decreased velocity resolution into account. For positions without significant emission, we summed the SNR channel maps centered at the mean velocity and using the mean line width of all consecutive pixels with SNR $> 2.5$ as integration borders (160 km s$^{-1}$). The spectral cube has an angular resolution of $\theta_x \times \theta_y = 0.225\arcsec \times 0.166\arcsec$ (spatial resolution of 1.55 kpc $\times$ 1.14 kpc at $z = 4.303$, $\theta_x$ and $\theta_y$ are the semi-major and semi-minor axes of the synthesized beam). It has a channel width of 13 km s$^{-1}$, and root mean square (RMS) noise $\sigma_\mathrm{[C\textsc{ii}]} = 0.5$ mJy beam$^{-1}$. 

\section{Methods} \label{sec:methods}

When studying galaxy kinematics in distant sources, the finite beam size of the observations is often larger than the spatial region over which a change in velocity takes place, in an effect known as beam-smearing \citep{Davies_2011, Swaters_2009}. The commonly-used tilted-ring modelling technique \citep[see][]{Rogstad_1974} has been used to extract galaxy kinematics in 2D \citep{Begeman_1989, van_der_Hulst_1992, Krajnovic_2006, Sellwood_2015}, and also in 3D \citep{Sicking_1997, Jozsa_2007, 3db_2015}. 2D methods require an intermediate step between the model and observations, where a velocity field must be extracted from the spectral cube. Several methods can be used to perform this step, and they can often be inconsistent amongst themselves \citep{3db_2015}. They also perform poorly with low angular resolution data due to beam-smearing. 3D methods have an advantage where they skip this intermediate stage to work directly with the spectral cube, naturally taking into account the effect of the synthesized beam to provide better results in kinematic analyses. In this work, we use \texttt{\textsuperscript{3D}BAROLO} (3D-Based Analysis of Rotating Object via Line Observations, \citealt{3db_2015}) for its ability to handle low SNR data and low angular resolution data.

    \subsection{Source detection using \texorpdfstring{\normalfont{\texttt{\textsuperscript{3D}BAROLO}}}{3Dbarolo}} \label{subsec:3Dsearch}

    Using the same naming convention presented in \citet{Hill_2020}, we extract $ 2\arcsec \times 2\arcsec $ ($ 3\arcsec \times 3\arcsec $ for source C1) sub-cubes for the sources (green boxes in \cref{fig:spt2349_mosaic}) using the \texttt{imsubimage} task in \texttt{CASA}. We note that sources C3 and C6 appear off-center due to their proximity to nearby galaxies. These sub-cubes are then processed with the \texttt{SEARCH} task in \texttt{\textsuperscript{3D}BAROLO}. The task detects sources in position-position-velocity space, creating a mask using a combination of smoothing and clipping. Here, we use only the clipping options, identifying the sources at 5$\sigma_{\mathrm{[C\textsc{ii}]}}$ and growing them down to 2.1$\sigma_{\mathrm{[C\textsc{ii}]}}$ around the emission peaks for sources C2--C7, and 2.5$\sigma_{\mathrm{[C\textsc{ii}]}}$ for source C1. The lower threshold includes sufficient signal without introducing excessive noise around the sources, while also being the SNR limit for use with \texttt{\textsuperscript{3D}BAROLO}. Using these thresholds, sources C1--C11 and C13 are identified in the [C\textsc{ii}] spectral cube. Sources C12 and C14--C23, identified in \citet{Hill_2020}, are not detected here (see discussion in \cref{sec:results}).

    \subsection{Kinematic fitting using \texorpdfstring{\normalfont{\texttt{\textsuperscript{3D}BAROLO}}}{3Dbarolo}} \label{subsec:3Dfitting}

    The \texttt{3DFIT} task requires initial values for an array of geometrical parameters such as the kinematic center ($x_0$, $y_0$), position angle ($\phi$), and inclination angle ($i$), along with kinematic parameters such as the rotation velocity ($V_{\mathrm{rot}}$), velocity dispersion ($\sigma_{\mathrm{disp}}$), and the systemic velocity ($V_{\mathrm{sys}}$) to perform a rotation curve fit in 3D-space. A model observation is built in 3D space as concentric rings based on the initial parameter values, allowing for the gas motion to be characterized in a disc as rotational or turbulent. The model is then convolved with the beam, before performing a $\chi^2$--minimization between the observed line emission and the model. The code continues to re-compute models until a best-fitting model is found. The advantage of such a fitting process is the reduction of the aforementioned beam-smearing effects, where the comparable sizes of the source and the synthesized beam could underestimate the rotation velocity and, at the same time, overestimate the gas velocity dispersion \citep{3db_2015}. We note that the term `velocity dispersion' refers to the galaxy's intrinsic value, corrected for instrumental effects of spatial and spectral resolution. The term `broadening' will refer to the non-corrected velocity dispersion. 

    We provide initial guesses for $x_0$, $y_0$, $V_{\mathrm{rot}}$ and $V_{\mathrm{sys}}$ using the results from the \texttt{SEARCH} task. The values of $x_0$, $y_0$ obtained as such are the flux-weighted average of the moment-0 map. We find that for some galaxies, $x_0$ and $y_0$ do not overlap with the kinematic center of the galaxy (where the line-of-sight velocity $V_{\mathrm{LOS}} \approx 0$ km s$^{-1}$). For these cases, we use the moment-1 (velocity field) map and the position-velocity ($p-v$) maps to adjust the coordinates before supplying them as initial guesses. $\phi$ is measured counter-clockwise from the north towards the receding side of the galaxy using the moment-1 map. For $i$, we first compute $(b/a)$, where $a$ and $b$ are the semi-major and semi-minor axes of the source, respectively, using the \texttt{CASA imageanalysis} tool \texttt{fitcomponents}. $i$ is then estimated using $ \cos^2 i = ((b/a)^2 - q^2) / (1-q^2)$, where $q = 0.2$ for a mildly thick disc. We note here that for galaxies that appear edge-on (C1, C4, and C5), we use $\cos i = (b/a)$ since the former definition consistently overestimates the inclination angle by $\approx 5\degr$. The estimated $i$ is also verified against a fraction-of-peak contour that encloses 20\% of the peak moment-0 intensity, $\mathrm{[C\textsc{ii}]_{peak}}$. Lastly, $\sigma_{\mathrm{disp}}$ is arbitrarily set at values between 30--150 km s$^{-1}$. 

    Other relevant parameters used in the analysis are the ring separation (\texttt{RADSEP}), number of rings (\texttt{NRADII}), and normalization type (\texttt{NORM}). Generally, \texttt{RADSEP} defaults to the major axis of the synthesized beam in order to describe rotation curves with independently estimated points along the rotation curve. In our case, we use \texttt{RADSEP} $= \sqrt{\theta_x \theta_y}/2$, achieving a close-to-Nyquist sampling\footnote{Nyquist sampling is when the synthesized beam is sampled $\approx2.4$ times.} of the rotation and dispersion curves. \texttt{NRADII} provides the number of rings that are spaced at a width of \texttt{RADSEP}, where the kinematic parameters are extracted. Lastly, \texttt{NORM} allows for choosing between a pixel-to-pixel or an azimuthal normalization of the moment-0 map (see \citealt{3db_2015} for more details). Here we choose azimuthal normalization. 

    We finally run the \texttt{3DFIT} task to simultaneously fit four parameters per ring -- $V_{\mathrm{rot}}$, $\sigma_{\mathrm{disp}}$, $\phi$ and $i$ -- using a two-stage fitting process. The code allows the user to choose the regularization type for the geometry. Here, we use a Bezier\footnote{\href{https://encyclopediaofmath.org/wiki/Bézier_curve}{Encyclopedia of Mathematics}} fitting of geometrical parameters in the first stage, while the code automatically fits the kinematic parameters in the second stage. $x_0$, $y_0$, and $V_{\mathrm{sys}}$ are kept constant throughout after verification using the ($p-v$) diagrams and the moment-1 map. We assume a thin-disc approximation and set the scale height $z_{0}$ to 1/10th of the half-light radius of the individual sources (see \citet{Hill_2020} for values), and discuss the implications in \cref{sec:discussion}. We also assume the large-scale radial motion of the gas is obscured in the high rotation velocities, and set the radial velocity $V_{\mathrm{rad}} = 0$. A sample parameter file used in the fitting is presented in Appendix \ref{sec:initial_values_and_parameter_file}.

    \begin{figure*}
        \epsscale{1}
        \plotone{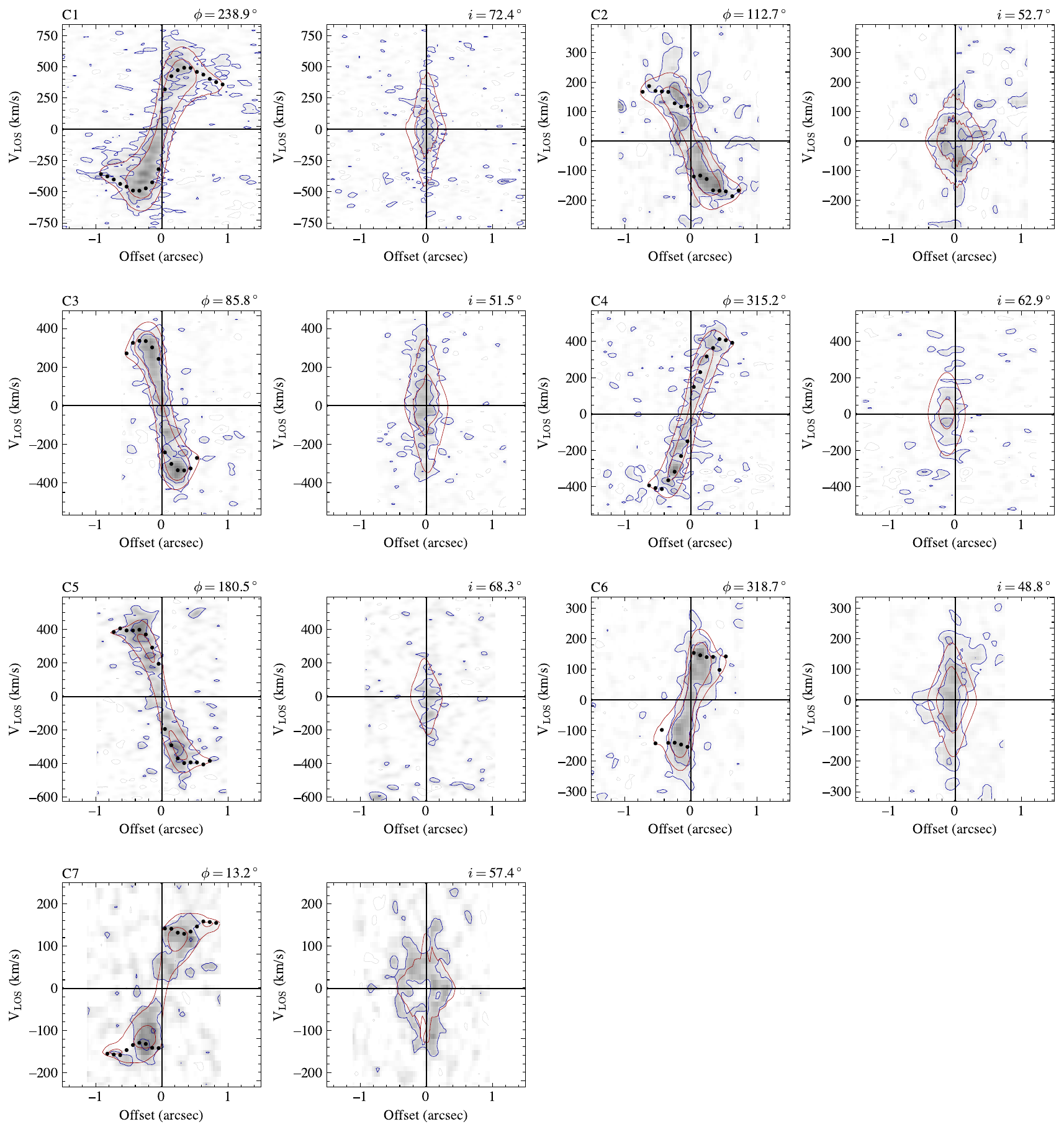}
        \caption{Position-velocity ($p-v$) plots for sources C1--C7 obtained from \texttt{\textsuperscript{3D}BAROLO}. Each slice is derived from un-masked cubes. In each $1 \times 2$ sub-figure, the left and right panels are the $p-v$ slices along the major and minor axes respectively. $\phi$ and $i$ are the derived position and inclination angle, respectively, for each source. The blue and red contours are at $n \sigma_{\mathrm{CII}}$ ($n = 2, 5$) for the data and model, respectively. The grey contours are at $-n \sigma_{\mathrm{CII}}$. The derived rotation curve is over-plotted with black filled circles.} \label{fig:spt2349_pvs_azim}
    \end{figure*}

    \begin{figure*}
        \epsscale{1}
        \plotone{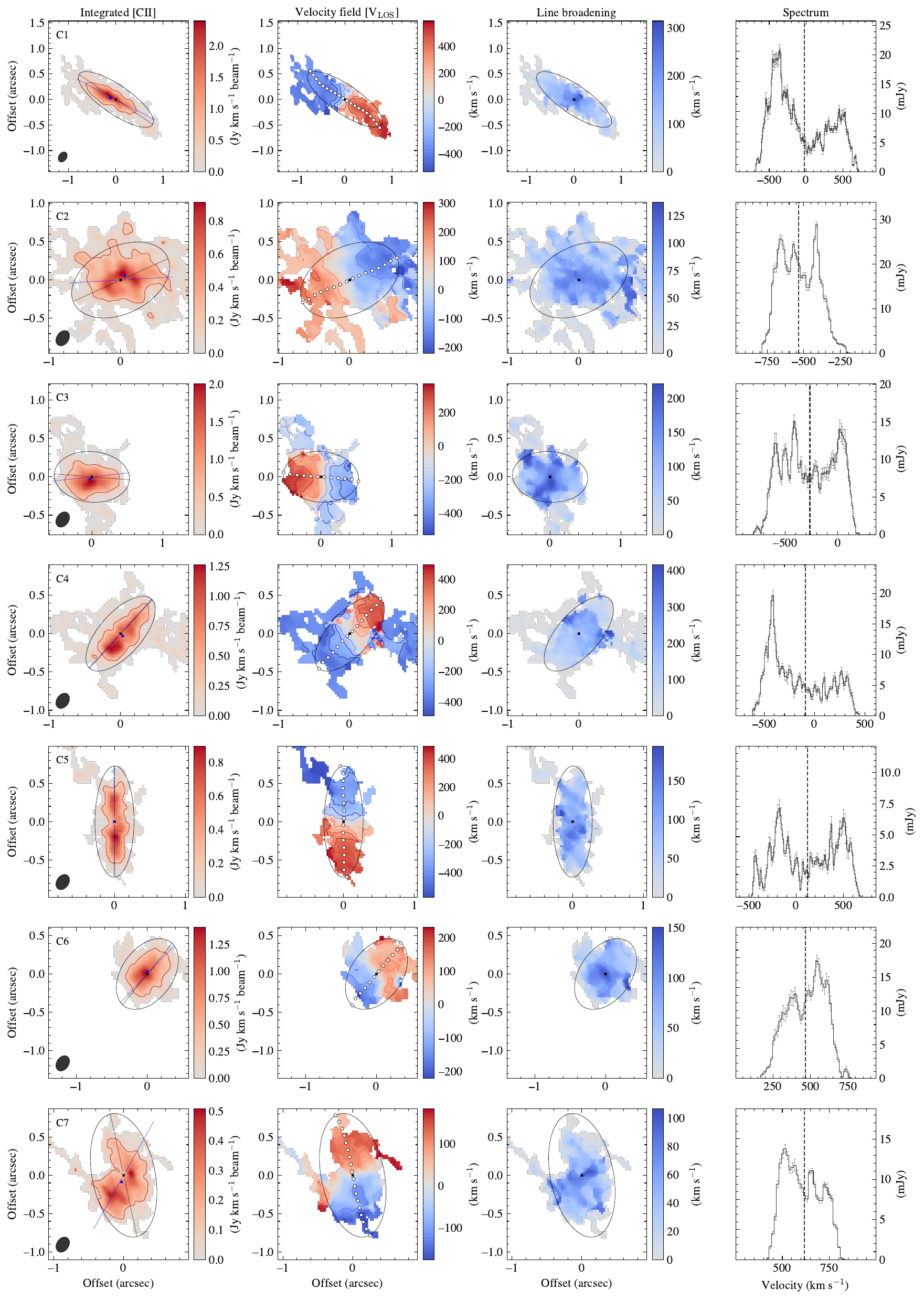}
        \caption{Moment maps and spectra from \texttt{\textsuperscript{3D}BAROLO} for sources C1--C7. The first, second, and third columns are the moment-0, moment-1, and moment-2 maps respectively. These moment maps are generated from data cubes that are not corrected for beam smearing. For each source, the synthesized beam is shown in the bottom-left of the moment-0 map. \textit{Column 1:} The black ellipse is the outer-most ring used in fitting with \texttt{\textsuperscript{3D}BAROLO}. The black and blue crosses are the kinematic, and flux-weighted centers of the [C\textsc{ii}] emission, respectively. The black and blue dashed lines are the kinematic and morphological major axes, respectively. The red contour is at $ 0.2\mathrm{[C\textsc{ii}]_{peak}}$. \textit{Column 2:} The filled white circles mark the kinematic major axis of the galaxy, respectively. \textit{Column 4:} Spectrum of the source obtained from a data cube masked according to the \texttt{SEARCH} task. The dashed vertical line is the systemic velocity of the source.} \label{fig:spt2349_kinmaps}
    \end{figure*}

    \begin{figure*}
        \epsscale{1}
        \plotone{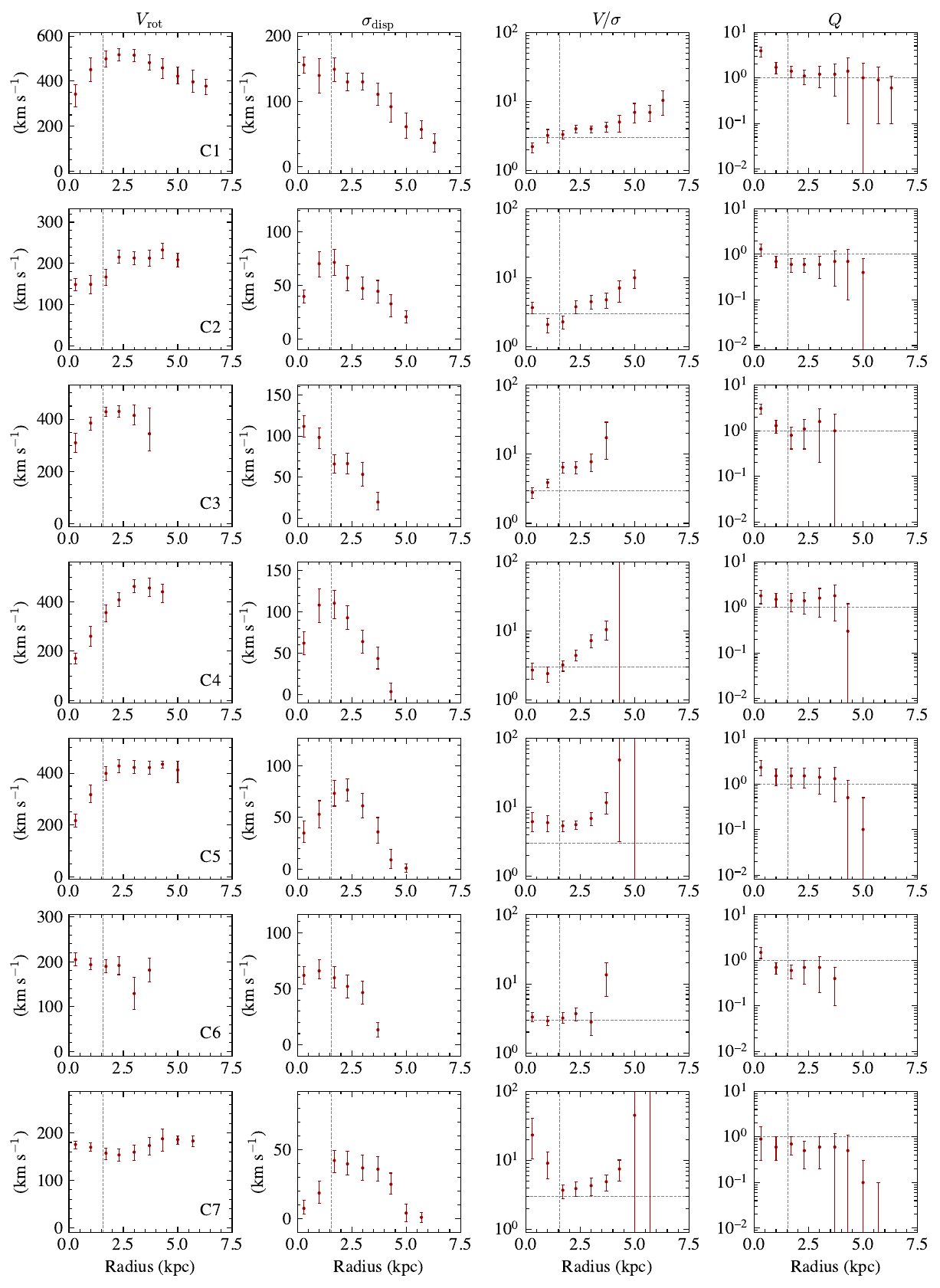}
        \caption{Results from \texttt{\textsuperscript{3D}BAROLO} for sources C1--C7. Columns 1 and 2 show the velocity and dispersion profiles, where the errors were obtained from a Markov Chain Monte Carlo sampling of the parameter space surrounding the best-fit values. Column 3 shows the $V_{\mathrm{rot}}/\sigma_{\mathrm{disp}}$ profile for each source. Column 4 shows the Toomre parameter profile for each source (see \cref{subsec:toomre_parameter_analysis}). The vertical dashed lines denote the beam extent ($\approx 1.6$ kpc). The horizontal dashed line in column 3 marks the demarcation line between rotation- and pressure-supported systems. The horizontal dashed line in column 4 marks the $Q = 1$ stability criteria.}
    \label{fig:spt2349_velocity_profiles}
    \end{figure*}

\section{Results} \label{sec:results}

At the end of the kinematic analysis, we find converging models for sources C1--C7, i.e., the seven FIR brightest sources listed in Table 3 of \citet{Hill_2020}. While we were able to detect sources C8--C11 and C13 using the \texttt{SEARCH} task, we find that due to their spatial extent approaching the synthesized beam size, too few data points exist to properly model their kinematics. Sources C12, and C14--C23 are not detected in the \texttt{SEARCH} task due to reduced sensitivity when compared to \citet{Hill_2020}. They report $\sigma_{\mathrm{[C\textsc{ii}]}} = 0.22$ mJy beam$^{-1}$ channel$^{-1}$ with a similar channel width, and a beam size that is 28\% larger than that for our sample. However, they use a combination of Cycle 5 and 6 ALMA data to produce a single, deep [C\textsc{ii}] data cube focusing on sensitivity, whereas we only use Cycle 6 data for its longer baseline, therefore prioritizing spatial resolution. Overall, this drop in sensitivity explains the lack of detection of the sources C12, and C14--C23. 

For each source, \cref{fig:spt2349_pvs_azim} and \cref{fig:spt2349_kinmaps} show the position-velocity ($p-v$) slices, the moment maps and the spectra. The $p-v$ slices are across the kinematic major and minor axes, while the moment maps and spectra are obtained from a spectral cube masked using \texttt{SEARCH} task parameters. We note that the $p-v$ slices and moment maps are not corrected for beam smearing. In \cref{tab:physical_params} and \cref{tab:kin_params}, we present the resulting kinematic center, position angle, inclination angle, and systemic velocity with respect to $z = 4.303$ (the flux-weighted kinematic center of the protocluster core).

    \subsection{Position-velocity slices} \label{subsec:position_velocity_slices}

    \textit{Major-axis slice:} For source C1, we find a rising and flattening of the rotation curve. For the remaining sources, the rotation curve rises, but cannot be measured out to distances where it is expected to flatten ($>3$ kpc). There is, however, a symmetry between the approaching and receding sides in all the sources, showing resemblance to the characteristic `S'-shape of a rotation-supported disc \citep{deBlok_2008}.\\
    \textit{Minor-axis slice}: At low angular resolution, a minor-axis $p-v$ slice of a moderately inclined rotating disc will present a broad, `blooming' pattern due to line broadening from instrumental effects \citep{3db_2015}. This behavior is present in all our sources. 

    \begin{deluxetable*}{CccCCCC}
        \tablewidth{\textwidth}
        \tablecaption{The morphological centers, position angles, and the derived geometrical parameters after kinematic modeling. The sources are named according to \citet{Hill_2020} and \citet{Miller_2018} in brackets. \label{tab:physical_params}}
        \tablehead{
        \colhead{Source} & \colhead{R.A., Dec. (morph)\tablenotemark{a}} & \colhead{R.A., Dec. (kin)\tablenotemark{b}} & \colhead{$\phi_{\mathrm{morph}}$\tablenotemark{c}} & \colhead{$\phi_{\mathrm{kin}}$\tablenotemark{d}} &\colhead{$i$\tablenotemark{e}} & \colhead{$z_0$\tablenotemark{f}}\\
        \colhead{} & \colhead{(J2000)} & \colhead{(J2000)} & \colhead{(\degr)} & \colhead{(\degr)} & \colhead{(\degr)} & \colhead{(kpc)}}
        \startdata
        \mathrm{C1\;(A)} & 23:49:42.7 $-$56:38:19.4 & 23:49:42.6 $-$56:38:19.4 & 244.3 \pm 0.4 & 238.9 \pm 2.4 & 72.4 \pm 1.1 & 0.2 \\
        \mathrm{C2\;(J)} & 23:49:43.2 $-$56:38:30.1 & 23:49:43.2 $-$56:38:30.1 &  93.1 \pm 5.0 & 112.7 \pm 1.3 & 52.7 \pm 1.2 & 0.2 \\
        \mathrm{C3\;(B)} & 23:49:42.8 $-$56:38:23.8 & 23:49:42.8 $-$56:38:23.8 &  95.7 \pm 2.1 &  85.8 \pm 1.2 & 51.5 \pm 0.2 & 0.1 \\
        \mathrm{C4\;(D)} & 23:49:41.4 $-$56:38:22.4 & 23:49:41.4 $-$56:38:22.4 & 316.0 \pm 1.3 & 315.2 \pm 1.3 & 62.9 \pm 0.7 & 0.1 \\
        \mathrm{C5\;(F)} & 23:49:42.1 $-$56:38:25.8 & 23:49:42.1 $-$56:38:25.8 & 180.6 \pm 0.9 & 180.5 \pm 2.7 & 68.3 \pm 1.5 & 0.2 \\
        \mathrm{C6\;(C)} & 23:49:42.8 $-$56:38:25.1 & 23:49:42.8 $-$56:38:25.1 & 314.5 \pm 2.5 & 318.7 \pm 1.3 & 48.8 \pm 1.4 & 0.1 \\
        \mathrm{C7\;(K)} & 23:49:43.0 $-$56:38:18.1 & 23:49:42.9 $-$56:38:18.0 & 328.5 \pm 5.7 &  13.2 \pm 2.0 & 57.4 \pm 1.2 & 0.2 \\
        \enddata
        \tablenotetext{a,b}{\hspace{10pt}Morphological and kinematic center of the source, respectively.}
        \tablenotetext{c,d}{\hspace{10pt}Morphological and derived kinematic position angles, respectively.}
        \tablenotetext{e}{Derived inclination angle.}
        \tablenotemark{f}{Scale height used in kinematic analysis, $0.1 R_{1/2}$ from \citet{Hill_2020}.}
    \end{deluxetable*}

    \vspace{-8mm}

    \subsection{Moment maps and spectra} \label{subsec:moment_maps}

    For our sample, we find from column 1 in \cref{fig:spt2349_kinmaps} that there is minimal scatter ($<0.22\arcsec$) between the kinematic center (black cross) and the flux-weighted average of the moment-0 map (blue cross). The red contour is at $0.2\mathrm{[C\textsc{ii}]_{peak}}$, and the black ellipse describes the outermost evaluated ring. There is good agreement between the two for all the sources, except in the case of C2 where the [C\textsc{ii}] emission appears to be asymmetrically concentrated around the center within a small region. In the case of source C7, we find sub-structures at this spatial scale that are perpendicular to its kinematic major axis. 
    
    The position angle for the morphological major axis is derived using the \texttt{CASA imageanalysis} tool \texttt{fitcomponents}, and is reported in \cref{tab:physical_params}. It is represented by the blue dashed line. We note that the angle obtained from the tool is degenerate by $180\degr$, and we report the adjusted value in \cref{tab:physical_params}. The difference between the morphological and kinematic axes (black dashed line in \cref{fig:spt2349_kinmaps}) is less than $30\degr$ for all the sources except C7.

    Column 2 in \cref{fig:spt2349_kinmaps} shows the moment-1 maps, obtained as the intensity-weighted velocity of the spectral line. In each source, we find a smooth velocity gradient, with the kinematic and morphological major axes broadly aligned with each other. The blue and red contours constitute a `spider' diagram for each source and are placed at $\pm n100$ ($n=1,2,3)$ km s$^{-1}$ \citep{Begeman_1989}. There are no closed contours along the major axis, suggesting that there is no peak in the rotation curve. The exceptions are C6 and C7, where the map shows a closed loop. Since there is no peak in the velocity profile (see \cref{fig:spt2349_velocity_profiles}, column 1), we suggest this is due to a combination of observing a compact disc, low SNR and remnant beam effects. The black cross is the kinematic center of the galaxy. 

    Column 3 in \cref{fig:spt2349_kinmaps} shows the line broadening map for the sources, i.e., a velocity dispersion map that is not corrected for instrumental broadening. We reiterate here that \cref{fig:spt2349_kinmaps} shows the moment maps obtained directly from the data cubes, without correcting for beam smearing. They are not used in the kinematic extraction and are presented here for completeness only. Appendix \ref{sec:model_residual_maps} shows the observed, modeled, and residual moment maps for sources C1--C7.

    Column 4 in \cref{fig:spt2349_kinmaps} shows the spectrum of each source. The dashed vertical line is the systemic velocity of the source. We find double-peaked profiles with varying degrees of assymmetry across the systemic velocity for all the sources.

    \subsection{Velocity profiles} \label{subsec:velocity_profiles}

    Columns 1 and 2 in \cref{fig:spt2349_velocity_profiles} show the rotation curve and velocity dispersion profiles respectively, derived using the spectral cubes of the sources. The errors are obtained via a Markov-Chain Monte Carlo (MCMC) sampling of the parameter space around the best-fit value. We note here that the errors are not statistical errors, but a measure of how the well-behaved the parameter space is around the best-fit value \citep{3db_2015}. In sources C3--C6, there is insufficient data to estimate the outermost parts of the rotation curve ($>3.5$ kpc). In \cref{tab:kin_params}, we tabulate the noise-weighted mean rotation velocity and dispersion, calculated by excluding the data points falling within the beam size ($<1.55$ kpc, vertical dashed line in \cref{fig:spt2349_velocity_profiles}) to avoid possible remnant beam-smearing effects. We are then left with at least three independent measurements for each velocity and dispersion profile. Since we have a reasonable estimate of the inclination angles, the computed $V_{\mathrm{rot}}$ values are robust. The estimated $\sigma_{\mathrm{disp}}$ for all the galaxies are greater than the velocity resolution of 13 km s$^{-1}$, and we can therefore expect the values produced by \texttt{3DFIT} to be reliable.

\section{Discussion} \label{sec:discussion}

\begin{deluxetable*}{CCCCCCCCCC}
    \tablecaption{The derived velocity and stability parameters after kinematic modeling. In the last row, we give the mean value for the sample (excluding source C7), where the error corresponds to the uncertainty of the mean, while the number in brackets gives the standard deviation of the sample. We note here that there is a slight discrepancy between the reported $(V_{\mathrm{rot}}/\sigma_{\mathrm{disp}})$ ratio to the value that is obtained when using $(V_{\mathrm{rot,mean}}/\sigma_{\mathrm{disp,mean}})$. This is simply due to the fact that for individual sources, $(V_{\mathrm{rot}}/\sigma_{\mathrm{disp}})_\mathrm{mean} \neq (V_{\mathrm{rot,mean}}/\sigma_{\mathrm{disp,mean}})$, ignoring the weights. \label{tab:kin_params}}
    \tablehead{
    \colhead{Source} & \colhead{$V_{\mathrm{sys}}$\tablenotemark{a}} & \colhead{$V_{\mathrm{rot}}$\tablenotemark{b}} & \colhead{$\sigma_{\mathrm{disp}}$\tablenotemark{c}} & \colhead{$(V_{\mathrm{rot}}/\sigma_{\mathrm{disp}})$\tablenotemark{d}} & \colhead{$Q_\mathrm{T}$\tablenotemark{e}} & \colhead{$f_{\mathrm{gas}}$\tablenotemark{f}}\\
    \colhead{} & \colhead{(kms$^{-1}$)} & \colhead{(kms$^{-1}$)} & \colhead{(kms$^{-1}$)} & \colhead{} & \colhead{} & \colhead{}
    }
    \startdata
    \mathrm{C1\;(A)} &  -21.5 & 472.3 \pm 12.0 & 95.6 \pm 5.5 & 4.0 \pm 0.2 & 1.1 \pm 0.2 &  0.35 \\
    \mathrm{C2\;(J)} & -534.5 & 209.3 \pm  7.3 & 37.6 \pm 3.8 & 3.3 \pm 0.4 & 0.6 \pm 0.1 &  0.20 \\
    \mathrm{C3\;(B)} & -265.1 & 425.4 \pm 12.4 & 49.3 \pm 6.0 & 6.7 \pm 0.8 & 0.9 \pm 0.3 & <0.69 \\
    \mathrm{C4\;(D)} &  -93.5 & 427.6 \pm 14.1 & 50.6 \pm 5.9 & 4.1 \pm 0.4 & 1.3 \pm 0.4 &  0.54 \\
    \mathrm{C5\;(F)} &  118.2 & 425.3 \pm  9.1 & 20.8 \pm 3.3 & 5.9 \pm 0.6 & 0.8 \pm 0.3 &  0.40 \\
    \mathrm{C6\;(C)} &  466.3 & 184.2 \pm 10.1 & 35.7 \pm 4.3 & 3.3 \pm 0.4 & 0.6 \pm 0.2 &  0.17 \\
    \mathrm{C7\;(K)} &  618.7 & 173.3 \pm  4.9 & 14.9 \pm 2.4 & 4.2 \pm 0.5 & 0.2 \pm 0.1 &  0.31 \\ 
    \tableline
    \multicolumn{2}{c}{Sample Mean} & 357.1 \pm 4.5\;(114.7) & 48.4 \pm 2.0\;(23.5) & 4.5 \pm 0.2\;(1.3) & 0.9 \pm 0.1\;(0.3) & \\
    \enddata
    \tablenotetext{a}{Systemic velocity, relative to $z = 4.303$}
    \tablenotetext{b,c,d,e}{\hspace{25pt}Noise-weighted mean and uncertainty for rotation velocity, velocity dispersion, $V/\sigma$ ratio, and Toomre parameter at $r>1.55$ kpc.}
    \tablenotetext{f}{Gas fraction, as presented in \citet{Rotermund_2021}.}
\end{deluxetable*}


\vspace{-8mm}

We have presented the results of a 3D kinematic analysis of the seven FIR brightest galaxies within the SPT2349$-$56 protocluster core. To interpret the results, we first take a look at the assumptions made in the fitting process. First, we note that any further improvement in the derived kinematics in the inner part of the discs can only be achieved with a combination of higher resolution and deeper data.

Next, the two available types of normalization -- localized and azimuthal -- serve different purposes \citep{3db_2015}. The former constructs non-axisymmetric models, used to identify untypical regions such as small-scale clumps, outflows or holes that could affect the overall velocity profile fit. This is presented via a model moment-0 map that is identical to that of the observations, i.e., the integrated spectrum in each pixel is the same for the model and the observations. In the latter case, an axisymmetric model is constructed such that the model flux is normalized to the azimuthally averaged flux of the data in each ring. This model is used to identify warps or large-scale clumps. We compared the output from both cases and find no large-scale structures in the galaxies, but the azimuthal normalization presented a better fit to the geometry. The data, model, and residual moment maps, residual $p-v$ maps, and fitting results for the four free parameters obtained from the azimuthal fitting are presented in Appendix \ref{sec:model_residual_maps}. We also found there is no significant, large-scale radial motion present in the galaxies by setting the radial velocity as a free parameter with an initial value of $V_{\mathrm{rad}} = 0$. We note that any possible radial velocity component much less than the rotation velocity would be lost due to resolution effects and low sensitivities.

The simultaneous modeling of geometry and kinematics often leads to unphysical gaps in the velocity and dispersion profiles. In our case, the \texttt{3DFIT} code uses the first minimization stage to fit a Bezier function to both $\phi$ and $i$. The next stage is used purely to fit the velocity curves, holding the geometry constant. We find that this two-stage minimization process is efficient compared to user-supplied fixed angles.

The weakest link in kinematic modeling is the inclination angle. For galaxies that appear face-on, there is no significant visual distinction between $i = 5\degr$ and $i = 20\degr$, but their estimated rotation curves differ drastically since $V_{\mathrm{rot}} \propto V_{\mathrm{obs}}/ \sin i$. For reference, $\sin 20\degr/\sin 5\degr \approx 4$. In such cases, the estimation of the minor-to-major ($b/a$) axis ratio has to be done with great precision. Here, we use a number of techniques as described in \cref{subsec:3Dfitting} to initially estimate $b/a$, and all the galaxies clearly present initial values greater than $40\degr$.

Our assumption of a thin disc with $z_0 \approx 0.1$ times the half-light radius also affects the estimation of the inclination angle. The mean of the derived inclination angles for our sample is $\approx$ 60\degr,corresponding to the mean inclination of a random sample of galaxies \citep{Romanowsky_2012}. The sample also presents clear rotation curves, suggesting that we are not dealing with face-on discs, where the assumption of a thin disc may wreck our interpretation. While the generated model reproduces the [C\textsc{ii}] emission well, it is still a simplified assumption, since at high redshift, turbulence within the disc would give rise to a higher velocity dispersion and a thick disc. \citet{Neeleman_2020} present a comparison between assumptions of an infinitely thin disc and a disc where the scale height is 0.15--1 times the scale radius at $z\approx 4.2$. Their results show a decrease in the estimated inclination angles by $\approx 7\degr$. Since we use $z_0 =$ 0.07--0.15 times the scale radius provided in \citet{Hill_2020}, there may be a possibility that we are overestimating the inclination angles. However, in the absence of additional information about the disc thickness, we report values obtained using a simplified thin-disc assumption.

    \subsection{Kinematic analysis} \label{subsec:v_sig_ratio}

    \begin{figure*}
        \epsscale{1.17}
        \plotone{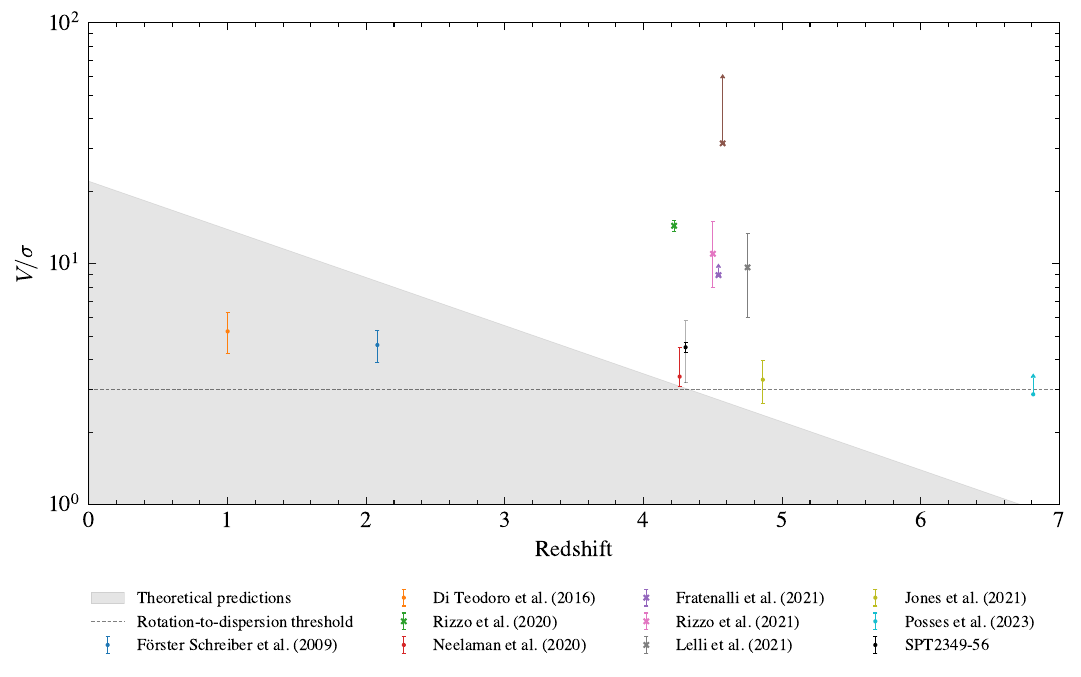}
        \caption{Comparison of the rotation-to-dispersion ratios of the galaxies within SPT2349$-$56 and samples of simulated and observed galaxies. The $V/\sigma$ ratio for our sample is obtained using $V_{\mathrm{flat}}$ and $\sigma_{\mathrm{ext}}$, both computed beyond $\approx 1.55$ kpc. The solid and faded black error bars are the uncertainty of the mean and standard deviation of the sample, respectively. The grey-shaded region includes the theoretical predictions made by \citet{Dekel_2014, Zolotov_2015, Hayward_2017} and \citet{Pillepich_2019}. The dashed line at $V/\sigma = 3$ marks the \citet{Burkert_2010} criteria we use for kinematic classification. Details of the comparison samples are presented in \cref{tab:comparison_Sample}. The colored dot and cross markers represent Main-Sequence (MS) and starburst galaxies respectively.} 
        \label{fig:spt2349_ratio_vs_redshift}
    \end{figure*}    

        \subsubsection{Rotators vs. Mergers} \label{subsubsec:rots_vs_merg}

        At the spatial resolution of our study, it remains difficult to distinguish between rotating discs and compact mergers. Several studies in the literature exist where the gas kinematics and morphology in FIR bright high redshift galaxies appear to be fully consistent with a rotating disc, but the authors favor the interpretation of a compact merger \citep[e.g.][]{Neri_2014, Litke_2019}. Merging of galaxies is a common interpretation for the cause of the extreme SFRs of local and high-redshift FIR bright sources. For the SPT2349-58 protocluster system, a strong interaction between the galaxies \citep{Rennehan_2020} is a likely trigger for the high SFRs in the gas-rich members, but as described in the following sections, we find evidence for rotating disc signatures along with mergers and/or tidal disturbance signatures in six of these galaxies.

        For a given system to be dynamically classified as rotation-supported, a series of criteria exist in the literature \citep{Foerster_Schreiber_2018, Wisnioski_2019, Foerster_schreiber_2020, Jones_2021}. The system:
        \begin{enumerate}
            \item Must have a monotonically smooth velocity gradient that defines the kinematic axis.
            \item Is rotation-supported and has $V/\sigma > 3$.
            \item Has spatially coincident morphological and kinematic centers.
            \item Has morphological and kinematic major axes separated by less than $30\degr$.
            \item Has a centrally-peaked velocity dispersion profile, such that the maximum defines the kinematic center.
        \end{enumerate}

       For our sample, we find that all galaxies satisfy condition 1. In this work, the morphological center of a galaxy is defined as the flux-weighted center of the moment-0 map (column 2 in \cref{tab:physical_params}). Therefore, for condition 3, if the distance between the kinematic and morphological center (determined using simple geometry) is less than $\theta_x = 0.22\arcsec$, then we assume that the centers are spatially coincident. The sources in our sample, except C7, satisfy condition 3.
        
        From \cref{subsec:moment_maps}, we have the difference between the kinematic and morphological axes to be $< 30\degr$ for all the sources except C7. In addition, if we use the definition for kinematic misalignment angle ($\phi_{\mathrm{mis}}$) from \citet{Franx_1991}, we find the morphological major axis of the [C\textsc{ii}] gas distribution co-aligned with the kinematic axis for the sample such that $\phi_{\mathrm{mis}} < 15\degr$. The exceptions are sources C2 and C7, where in the former, the emission appears to be centrally clumpy (see \cref{fig:spt2349_kinmaps}, columns 1 and 2), and in the latter, the sub-structures appear perpendicular to the kinematic axis. 
        
        Next, from \cref{fig:spt2349_velocity_profiles} we find that for all of our sources, excluding the points at $<0.22\arcsec$, the velocity dispersion profiles are centrally peaked. The dispersion maps presented in \cref{fig:spt2349_kinmaps} are broadening maps, and the increase in dispersion towards the edges is a result of low SNR and/or the broadening effect itself.

        In addition to these criteria, we refer to \cref{subsec:position_velocity_slices}, where we identified the presence of the distinctive symmetrical $p-v$ slices. Lastly, we refer to column 4 in \cref{fig:spt2349_kinmaps} where we find double-peaked profiles for all the sources, considered a signature of rotating discs \citep{Kohandel_2019}. Therefore, based on the conventional dynamical-support criteria, we find no immediate evidence supporting the presence of mergers in the protocluster environment, allowing us to further investigate gas kinematics. The exception is source C7, which appears to be a major merger, and we discard from the analysis in \cref{subsubsec:SPT2349}.

        \begin{deluxetable*}{cccCccC}
            \tablewidth{\textwidth}
            \tablecaption{Comparison samples with inclination-corrected rotation velocities and velocity dispersion values from the literature. We use the following notations:\\ 
            -- $V_{\mathrm{rot}}$ and $\sigma_{\mathrm{disp}}$ are the rotation velocity and velocity dispersion respectively, across all resolution elements.\\
            -- $V_{\mathrm{rot, flat}}$ and $\sigma_{\mathrm{disp, ext}}$ are the rotation velocity and velocity dispersion respectively, computed beyond/at the specified radius.\\
            \label{tab:comparison_Sample}}
            \tablehead{
            Study\tablenotemark{a} & Type of Objects & Tracer & $z$ & $V$ & $\sigma$ & $V/\sigma$ 
            }
            \startdata
            \citet{Foerster_Schreiber_2009} & Main-Sequence                 & H$\alpha$      & 2.08 & $V_{\mathrm{rot}}$ & $\sigma_{\mathrm{disp}}$ & $4.6 \pm 0.6$ \\ 
            \citet{Teodoro_2016}            & Main-Sequence                 & H$\alpha$      & 0.90 & $V_{\mathrm{flat}}$ & $\sigma_{\mathrm{ext}}$ & $5.5 \pm 0.3$ \\ 
            \citet{Rizzo_2020}              & Starburst                     & [C\textsc{ii}] & 4.22 & $V_{\mathrm{flat}}$ & $\sigma_{\mathrm{ext}}$ & $ 14.4 \pm 0.4$ \\ 
            \citet{Neeleman_2020}           & Main-Sequence                 & [C\textsc{ii}] & 4.26 & $V_{\mathrm{rot}}$ & $\sigma_{\mathrm{disp}}$ & $ 3.4_{-0.3}^{+1.1}$ \\ 
            \citet{Fratenalli_2021}         & Starburst                     & [C\textsc{ii}] & 4.54 & $V_{\mathrm{flat}}$ & $\sigma_{\mathrm{ext0}}$ & $ 9.0 \pm 0.7$\\ 
            \citet{Rizzo_2021}              & Starburst and Main-Sequence   & [C\textsc{ii}] & 4.50 & $V_{\mathrm{flat}}$ & $\sigma_{\mathrm{ext}}$ & $ 11 \pm 4$\\ 
            \citet{Lelli_2021}              & Starburst                     & [C\textsc{ii}] & 4.75 & $V_{\mathrm{rot}}$ & $\sigma_{\mathrm{disp}}$ & $9.7 \pm 3.7$ \\ %
            \citet{Jones_2021}              & Main-Sequence                 & [C\textsc{ii}] & 4.86 & $V_{\mathrm{rot}}$ & $\sigma_{\mathrm{disp}}$ & $3.3 \pm 0.7$ \\
            \citet{Posses_2023}             & Main-Sequence                 & [C\textsc{ii}] & 6.81 & $V_{\mathrm{flat}}$ & $\sigma_{\mathrm{ext}}$ & $ 2.9 \pm 0.5$\\
            \enddata
            \tablenotetext{}{\textit{Note-} The samples are shown in \cref{fig:spt2349_ratio_vs_redshift}.}
            \tablenotetext{a}{The studies chosen depend on two criteria - (i) It uses 3D kinematic modeling methods, (ii) It uses an ionized tracer.}
        \end{deluxetable*}
        
        \vspace{-8mm}

        \subsubsection{Gas kinematics} \label{subsubsec:gas_kinematics}
        
        Gas kinematics can help study the measure of rotational support in a galaxy. In this context, the rotational velocity and velocity dispersion correspond to \textit{ordered} and \textit{random} motions, respectively. There are a few definitions available in the literature to quantify this, such as $\Delta V_{\mathrm{obs}}/2\sigma_{\mathrm{int}}$ and $V_{\mathrm{rot}}/\sigma_{\mathrm{disp}}$ \citep{Foerster_Schreiber_2009, Rizzo_2020}. Here, $V_{\mathrm{obs}}$ is the velocity gradient of the tracer, and $\sigma_{\mathrm{int}}$ is the source-integrated velocity dispersion, where both are uncorrected for inclination. $V_{\mathrm{rot}}$ and $\sigma_{\mathrm{disp}}$ are the inclination-corrected rotation velocity and velocity dispersion respectively, where two further definitions can be derived. The first is obtained by setting $V_{\mathrm{rot}} = V_{\mathrm{max}}$, the maximum inclination-corrected rotation velocity and $\sigma_{\mathrm{disp}} = \sigma_{\mathrm{mean}}$, the mean velocity dispersion of the profile. The second definition is used for our sample, where we set $V_{\mathrm{rot}} = V_{\mathrm{flat}}$ and $\sigma_{\mathrm{disp}} = \sigma_{\mathrm{extended}}$, both measured beyond $0.22\arcsec$ ($\approx 1.55$ kpc, the beam size, marked as a vertical dashed line in \cref{fig:spt2349_velocity_profiles}). 
        
       We exclude the innermost points due to possible remnant beam-smearing effects in the central regions. We use the \citet{Burkert_2010} definition, where a galaxy is dynamically hot or pressure-supported if $V/\sigma \le 3$, and dynamically cold or rotation-supported if $V/\sigma > 3$. We find, however, that there are other thresholds used across different works (see \citet{Turner_2017} and references therein).

        Simulations and comparative observational studies have found that the velocity dispersion measured from ionized tracers tends to be $\approx$ 15 km s$^{-1}$ higher than those obtained from molecular or neutral tracers, implying a lower $V/\sigma$ when using ionized tracers \citep{Ubler_2019, Kretschmer_2022}. Similar to recent high-redshift studies that use ionized gas as a tracer to study gas kinematics, we use the [C\textsc{ii}] 158-$\micron$ line to study the galaxies within the SPT2349$-$56 protocluster core.

        \subsubsection{\texorpdfstring{SPT2349$-$56}{SPT2349-56}} \label{subsubsec:SPT2349}

        For the galaxies in our sample, we observe a rise and flattening of the rotation curves in sources C1--C5, and only flattening in source C6, while the dispersion profiles are all centrally peaked (excluding the beam, $<1.55$ kpc) and decaying (see \cref{fig:spt2349_velocity_profiles}, columns 1 and 2 respectively). Column 3 in \cref{fig:spt2349_velocity_profiles} shows the complete $V/\sigma$ ratio profile for the galaxies in our sample, while Column 5 in \cref{tab:kin_params} shows the corresponding mean $V/\sigma$ ratio. All seven sources show $V/\sigma > 3$, categorizing them as rotation-dominated systems, also satisfying condition 2 in \cref{subsubsec:rots_vs_merg}. The mean of the sample (excluding source C7) is $4.5 \pm 0.2$. Since [C\textsc{ii}] traces the cold inner molecular disc, our sample, with its high $V/\sigma$ values, is in line with the results presented for the inner molecular disc in \citet{Kretschmer_2022} and \citet{Kohandel_2023}. \Cref{fig:spt2349_ratio_vs_redshift} summarizes the current understanding of the evolution for the $V/\sigma$ ratio with $z$, for both observational data and theoretical models. We discuss the relevant theoretical models and observational samples in the following sections. 

        \subsubsection{Numerical and analytical studies} \label{subsubsec:numerical_and_analytical_studies}
	
       The grey shaded region in \cref{fig:spt2349_ratio_vs_redshift} includes results and analyses from \citet{Dekel_2014, Zolotov_2015, Hayward_2017} and \citet{Pillepich_2019}. Through theoretical methods and cosmological simulations, \citet{Dekel_2014} and \citet{Zolotov_2015} show that beyond $z = 3$, galaxies are dominated by violent disc instabilities, with $V/\sigma \le 2$.
	
       \citet{Hayward_2017} presented an analytical model to study the consequences of stellar feedback processes. In this model, the velocity dispersion is not a measured quantity, but a derived one, and is obtained as $\sigma_{\mathrm{disp}} \approx f_{\mathrm{gas}} (V_\mathrm{c}/\sqrt{2})$\footnote{equation 20 in their paper, setting $Q_{\mathrm{turb}} = 1$}. Here, $f_{\mathrm{gas}} = M_{\mathrm{gas}}/(M_{\mathrm{gas}}+M_{\mathrm{star}})$ is the gas fraction, and $V_\mathrm{c} = \sqrt{V_\mathrm{gas}^2 + V_\mathrm{star}^2 + V_\mathrm{DM}^2}$ is the circular velocity. Since we do not have stellar and dark matter velocity contributions on hand, we substitute $V_\mathrm{c} = V_\mathrm{gas}$ and first set $f_{\mathrm{gas}} = 0.7$ \citep{Tadaki_2019, Rennehan_2020}. This leaves us with predicted $\sigma_{\mathrm{disp}}$ values that are approximately 2--5 times higher than the derived values. However, if we use the $f_{\mathrm{gas}}$ values presented in \citealt{Rotermund_2021} (column 7 in \cref{tab:kin_params}), we find that the predicted $\sigma_{\mathrm{disp}}$ values are between 0.5--3 times the derived values. This difference in predicted $\sigma_{\mathrm{disp}}$ values using the \citet{Hayward_2017} model possibly rises from the assumption that for high-redshift galaxies, the inferred gas fractions are usually high ($f_{\mathrm{gas}} \approx 0.7 $), spread over a range of 0.4--0.9 \citep{Tadaki_2019}.

       Lastly, \citet{Pillepich_2019} predict the evolution of $V/\sigma$ with respect to redshift ($0 \le z \le 5$) for a gas mass range of $10^9 - 10^{11}$ $M_\sun$ from the TNG50 simulation. At $z = 4.303$, we find that the mean SPT2349$-$56 value ($4.5 \pm 0.2$) is greater than the median TNG50 value of $3\pm 1.5$ by one standard deviation. We also note that in \cref{fig:spt2349_ratio_vs_redshift}, this evolution is extrapolated beyond $z = 5$.

        \subsubsection{Observational samples} \label{subsubsec:observation_samples}

        \Cref{tab:comparison_Sample} shows the observational samples used in \cref{fig:spt2349_ratio_vs_redshift} as comparisons, including the tracer, the definition of $V_{\mathrm{rot}}$ and $\sigma_{\mathrm{disp}}$, and the $V/\sigma$ values. We choose the samples based on two criteria - (i) The study uses an ionized gas tracer to study kinematics; (ii) The study uses a 3D kinematic modeling method to evaluate the kinematics. The objects in these studies are starbursts or Main-Sequence (MS) galaxies. Beyond $z = 2.5$, several samples deviate from the predicted $V/\sigma$ evolution. Starbursts (colored crosses in \cref{fig:spt2349_ratio_vs_redshift}) appear to have higher $V/\sigma$ compared to the MS samples (colored dots in \cref{fig:spt2349_ratio_vs_redshift}). SPT2349$-$56 consists mainly of MS galaxies \citep{Rotermund_2021, Hill_2022}, and its mean and corresponding error (black filled circle in \cref{fig:spt2349_ratio_vs_redshift}) is within the region between starbursts and MS samples at a similar redshift. 
        
        The comparison samples at a similar redshift are \textit{field} galaxies, i.e., they are not associated with a protocluster-like environment. In these cases, it is possible to achieve a rotating disc via `cold' processes (see \cref{sec:introduction}). However, if we assume that the SPT2349$-$56 core accretes via a `cold stream in a hot medium' as presented in \citet{Dekel_2009a}, we find that the tidal interaction timescale in a protocluster system is much shorter than the cold stream accretion time, rendering the cold streaming process inefficient. This is further discussed in \cref{sec:rotating_discs_in_a_protocluster_core}. 

    \subsection{Toomre parameter analysis} \label{subsec:toomre_parameter_analysis}

    In the absence of sufficient random motion in a thin, smoothly distributed rotating disc, the disc is expected to suffer violent instabilities \citep{Toomre_1964}. The Toomre parameter ($Q_\mathrm{T}$) measures the stability of a disc against gravitational perturbations. $Q_\mathrm{T}$ is defined as

    \begin{equation}\label{eq:toomre_q}
        Q_\mathrm{T} = \frac{\sigma_{\mathrm{disp}}\kappa}{\pi G \Sigma_{\mathrm{gas}}}\;,
    \end{equation}

    where $Q_\mathrm{T} < 1$ describes an unstable disc. $\kappa = \sqrt{R( \mathrm{d} \Omega^{2}/\mathrm{d} R) + 4\Omega^2}$ is the epicyclic frequency, with $\Omega = V_{\mathrm{rot}}/R$ (which simplifies to $\kappa = \sqrt{2}V_{\mathrm{rot}}/R$). $\Sigma_{\mathrm{gas}}$ is the gas mass surface density, which is obtained in our case as 

    \begin{equation} \label{eq:surface_density}
        \Sigma_{\mathrm{gas}} = \frac{L_{\mathrm{CII}}\alpha_{\mathrm{CII}}}{\mathrm{per\,unit\,area}}\quad(\mathrm{M_\sun\,kpc}^{-2})\;,
    \end{equation}

    where $L_{\mathrm{CII}}$ and $\alpha_{\mathrm{CII}}$ are the [C\textsc{ii}] line luminosity and the atomic gas mass to [C\textsc{ii}] line luminosity ratio, respectively. The $\alpha_{\mathrm{CII}}$ parameter takes a range of values between 5 -- 30 (Sulzenauer et al. (in prep.), \citealp{Zanella_2018, Rizzo_2020, Vizgan_2022}). Here we use values presented in \citet{Rizzo_2020}, where $\alpha_{\mathrm{CII}} = 7.3 \pm 1.2$ $M_\sun/L_\sun$. There is a possibility to use dynamical mass estimates to deduce $\alpha_{\mathrm{[C\textsc{ii}]}}$ for this sample, however, it is beyond the scope of this work.

    $L_{\mathrm{CII}}$ is found via replacing the CO line luminosity calculations with the [C\textsc{ii}] line in Eq. 1 of \citet{Solomon_1997}

    \begin{equation}
        L_{\mathrm{CII}} = 1.04 \times 10^{-3} S_{\mathrm{CII}}\Delta V \nu_{\mathrm{rest}}(1+z)^{-1} D_\mathrm{L}^2 \quad (\mathrm{L_\sun}),
    \end{equation}

    where $S_{\mathrm{CII}}\Delta V$ is the velocity integrated flux in Jy km s$^{-1}$, $\nu_{\mathrm{rest}}$ is the [C\textsc{ii}] line rest frequency in GHz, and $D_\mathrm{L}$ is the luminosity distance in Mpc.

    Studies such as \citet{Rizzo_2020} and \citet{Neeleman_2020} obtain mean values of $\approx 0.96$ beyond 1 kpc radii at redshift $\approx 4.2$. For our sample of galaxies, the mean value is $Q_\mathrm{T} = 0.8 \pm 0.1$, with individual values ranging from 0.2--1.1. The largest source of uncertainty is the gas surface density, which could be mitigated using sophisticated means of flux profile extraction in future works. Sources C2 and C7 show sub-structures in the moment-0 maps with low $Q_\mathrm{T}$ values, suggesting an imminent collapse of unstable discs. Source C6, found in the central region of the core, appears to be extremely compact in its moment-0 map with low $Q_\mathrm{T}$ values, suggesting unresolved sub-structuring at our spatial resolution. 
    
    \begin{figure}[h]
        \epsscale{1.17}
        \plotone{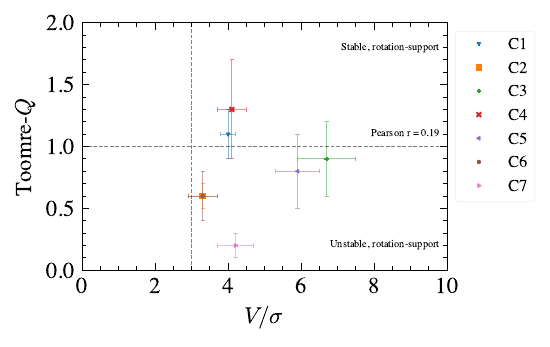}
        \caption{The Toomre parameter plotted against the $V /\sigma$ ratio. The dashed line at $Q_\mathrm{T} = 1$ marks the stability criteria, while the dashed line at $V/\sigma = 3$ marks the rotation-to-dispersion transition. Note that the markers for sources C2 and C6 overlap.} \label{fig:spt2349_q_vs_ratio}
    \end{figure}

    \begin{deluxetable*}{cCCCc}
        \tablewidth{\textwidth}
        \tablecaption{PVsplit parameter values derived for SPT2349$-$56.}
        \tablehead{\colhead{Source} & $P_{\mathrm{major}}$ & $P_{\mathrm{V}}$ & $P_{\mathrm{R}}$ & \colhead{Classification}\\
        \colhead{} & \colhead{Discs $\le 0 < $ Mergers} & \colhead{Discs $ < 0 <$ Mergers} & \colhead{Mergers $0 \leftrightarrow 1$ Discs} & \colhead{}
        \label{tab:PVsplit}}
        \startdata
        C1 (A) & -1.47 & -1.18 & 0.30 & Disturbed disc\\ 
        C2 (J) & -1.62 & -0.85 & 0.16 & Disturbed disc\\ 
        C3 (B) & -2.64 & -0.52 & 0.41 & Disc\\
        C4 (D) & -1.32 & -0.62 & 0.35 & Disturbed disc\\
        C5 (F) & -1.32 & -0.55 & 0.42 & Disturbed disc\\
        C6 (C) & -2.33 & -0.50 & 0.19 & Disturbed disc\\
        C7 (K) & -1.63 & -0.40 & 0.35 & Disturbed disc\\
        \enddata
    \end{deluxetable*}

    \vspace{-8mm}
     
    \Cref{fig:spt2349_q_vs_ratio} shows disc stability as a function of the $V /\sigma$ ratio. Sources C1 and C4 show rotation support with a stable disc, while the rest of the sources show rotation support with unstable discs. There is a mild positive correlation between disc stability with respect to dynamical support, with a Pearson-$r$ coefficient of $\approx 0.19$. In order to further explore the nature of these sources using their morphology, we utilize a newly developed tool, \texttt{PVsplit} \citep{Rizzo_2022}.

   \subsection{\texorpdfstring{\normalfont{\texttt{PVsplit}}}{PVsplit}} \label{subsec:PVsplit}

   \begin{figure}[b]
        \epsscale{1.1}
        \plotone{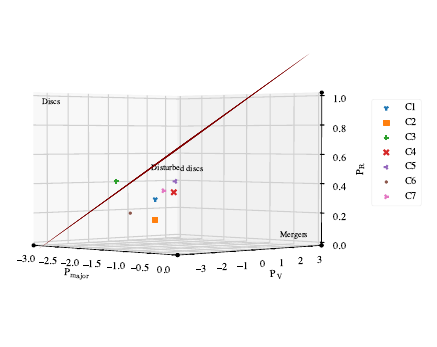}
        \caption{Results of the PVsplit tool. The 2D plane separates the mergers and discs. A 3D representation of this plot, with the same elevated viewing angle but different azimuthal viewing angles ($0\degr-360\degr$), is available online.}
        \label{fig:spt2349_pvsplit}
    \end{figure}
    
    In light of low-resolution data and its role in the incorrect kinematic classification of galaxies at high-redshift, \citet{Rizzo_2022} have developed a new tool, \texttt{PVsplit}. This method utilizes $p-v$ maps instead of moment maps to quantify the morpho-kinematic nature of galaxies using three empirical parameters, provided that there are at least three independent elements and SNR $ \ge 10$ across all the elements. The three parameters are defined as follows:

    \begin{itemize}
        \item[-] $P_{\mathrm{major}}$ : asymmetry of the major-axis slice with respect to the systemic velocity of the source; Discs $\le 0 < $ Mergers
        \item[-] $P_{\mathrm{V}}$ : distribution of the emission peak along the velocity axis; Discs $ < 0 <$ Mergers
        \item[-] $P_{\mathrm{R}}$ : distribution of the emission peak along the position axis; Discs: closer to 1, Mergers: closer to 0. 
    \end{itemize}
    
    Applying Eqns. 6, 7 and 8 in \citet{Rizzo_2022} to our $p-v$ maps, we present the results for our sample in \cref{tab:PVsplit} and \cref{fig:spt2349_pvsplit}. Targets in our sample do not meet the criteria for SNR, but we nevertheless apply the method since it gives valuable insights in the dynamical classification of the systems. The dividing plane in \cref{fig:spt2349_pvsplit} is obtained using $ -0.63P_{\mathrm{major}} - 0.27P_{\mathrm{V}} + 2.78P_{\mathrm{R}} - 2.72 = 0$ \citep{Roman_Oliviera_2023}. 
    
    Comparing \cref{fig:spt2349_pvsplit} with Figure 17 in \citet{Rizzo_2022}, it is immediately apparent that our sample occupies the same space as those of \textit{disturbed} discs and mergers. Based on the definitions of $P_{\mathrm{major}}$ and $P_{\mathrm{V}}$, we find from \cref{tab:PVsplit} that all the sources satisfy conditions for being `disc-like'. This implies two things - first, the sources are symmetric about the systemic velocity; second, the emission is either slightly spread out along the line-of-sight velocities or concentrated closer to the systemic velocity. Lastly, based on the intensity distribution along the position axis ($P_{\mathrm{R}}$) and the best-dividing plane, we find that only source C3 identifies as a disc in our sample.
    
    A possible explanation for the differences in classifications obtained via dynamical analysis (\cref{subsubsec:SPT2349}) and \texttt{PVSplit} is tidal torques. In a highly-overdense environement like that of a protocluster core, tidal torques can induce a velocity gradient across the galaxy, along with an asymmetric gas accumulation \citep{Gullberg_2019, MartinezGarcia_2023}. From column 4 in \cref{fig:spt2349_kinmaps}, we know that the sources display double-peaked spectra with varying degrees of asymmetry, arising from the corresponding aymmetric gas distributions. This interpretation also supports the low Toomre-$Q$ values obtained in \cref{subsec:toomre_parameter_analysis}. We therefore find that at this resolution and sensitivity, our sample adheres to the `disturbed discs' region, despite the high $V/\sigma$ values, i.e., they present with strong kinematic signatures and moderate morphological disturbances.

    \subsection{Rotating discs in a protocluster core} \label{sec:rotating_discs_in_a_protocluster_core}

    The number of observations describing the presence of cold rotating disc galaxies at high-redshift \citep{Pillepich_2019, Rizzo_2020, Neeleman_2020, Fratenalli_2021} is already challenging our current theories of galaxy evolution. In our case, we have found six seemingly well-ordered discs packed in an extreme environment containing 23 galaxies with average velocity $v = 360$ km s$^{-1}$ \citep{Hill_2020} within a radius of $\approx 100$ kpc. At $z = 4.303$, the age of the Universe is 1.4 Gyr. From \citet{Murdin_2001}, the crossing time for the protocluster core is 
    \begin{equation*}
        t_{\mathrm{cr}} = \frac{R}{v} = \frac{100\,\mathrm{kpc}}{360\,\mathrm{km\,s^{-1}}} = 0.28\,\mathrm{Gyr},
    \end{equation*}

    while the relaxation time is

    \begin{equation*}
        t_{\mathrm{r}} = t_{\mathrm{cr}} \frac{N}{8\ln{N}} = 0.27\frac{23}{8\ln{23}} = 0.25\,\mathrm{Gyr}.
    \end{equation*}

    In a protocluster system that is around 1 Gyr old (inferred from the stellar population, see \citealp{Rotermund_2021}), we have $t_{\mathrm{cr}} \approx t_{\mathrm{r}}$, meaning that the galaxies are losing energy via collisions every 0.27 Gyr; i.e., they very likely had a close interaction within their lifetime, suggesting that their gaseous discs should be disturbed, clearly evident in \cref{fig:spt2349_pvsplit}. 
    
   Explanations other than that of tidal torques in an overdense environement for ordered motion are that the discs are able to immediately recover after an interaction, or that the discs are simply so massive, that a strong interaction does not affect its stability, or that the classical definitions strongly depend on the SNR of the source.
    
    Numerical simulations of the SPT2349$-$56 protocluster system provide further evidence that the galaxies are undergoing strong interactions at a timescale of $\approx 0.5$ Gyr \citep{Rennehan_2020}. However, \citet{Remus_2022} present results from the \textit{Magneticum} simulation, wherein they reproduce the fast-rotating galaxies at the protocluster core, suggesting that the discs need not be completely destroyed even if they experience close interactions. This is also supported by recent observations of interacting galaxies in group environments at high-redshift, where the kinematics traced by [C\textsc{ii}] observations show well-ordered rotating discs, albeit in less dense environments \citep{Oteo_2016, Roman_Oliviera_2023}. Despite this possible evidence of another galaxy group presenting with ordered discs in its environment, we lack evidence to state whether the phenomenon of rotating discs in such environments is an outlier or the norm. 

\section{Conclusion} \label{sec:conc}

In this paper, we have presented kiloparsec-scale resolution ALMA observations of the [C\textsc{ii}] 158-$\micron$ transition of galaxies within the SPT2349$-$56 protocluster system at $z \approx 4.3$. Previous works describe the identification, multi-wavelength studies, and forward simulations of the system \citep{Miller_2018, Hill_2020, Rennehan_2020, Rotermund_2021, Hill_2022}, and here we report the kinematics of the constituent galaxies. The results are summarized as follows:

\begin{enumerate}
    \item The [C\textsc{ii}] emission of our sample of the seven submillimeter-bright galaxies is reproduced well using differentially-rotating models with a thin-disc assumption.
    \item The mean inclination-corrected rotation velocity of the sample is $V_{\mathrm{rot}} = 357.1 \pm 114.7$ km s$^{-1}$, and individual values between 170--475 km s$^{-1}$. 
    \item The mean intrinsic velocity dispersion of the sample is $48.4 \pm 23.5$ km s$^{-1}$, and individual values between 10--100 km s$^{-1}$. 
    \item On the basis of criteria found across literature, we find evidence for six rotating discs and one possible major merger at $z = 4.303$ within a protocluster environment. 
    \item The mean Toomre parameter for the sample is $Q_\mathrm{T} = 0.9 \pm 0.3$, where sources C1 and C4 are stable rotating systems, while the remaining sources classify as unstable rotating discs.
    \item Analyzing the $p-v$ maps to further study the morpho-kinematic nature of these sources, we find evidence for disc disturbance in regions $<10$ kpc, supporting the low Toomre-$Q$ values and high $V/\sigma$ values. That is, the sources are observed as disturbed rotating discs characterized by an asymmetric gas distribution.
\end{enumerate}

Overall, we find that the dynamical properties of the galaxies in our sample show considerable similarity to a number of other observations at a similar redshift. The main difference is that of morphology and the environment - while the comparison samples are not associated with turbulent environments, our galaxies exist in a highly active protocluster system that in itself has several unique properties. 

\section*{Acknowledgements}\label{sec:acknowledgements}

The authors would like to thank E. M. Teodoro for the many useful discussions about using \texttt{\textsuperscript{3D}BAROLO}. M. A. acknowledges support from FONDECYT grant 1211951, ANID+PCI+INSTITUTO MAX PLANCK DE ASTRONOMIA MPG 190030, ANID+PCI+REDES 190194 and ANID BASAL project FB210003. This paper makes use of the following ALMA data: ADS/JAO.ALMA\#2018.1.00058.S. ALMA is a partnership of ESO (representing its member states), NSF (USA) and NINS (Japan), together with NRC (Canada), MOST and ASIAA (Taiwan), and KASI (Republic of Korea), in cooperation with the Republic of Chile. The Joint ALMA Observatory is operated by ESO, AUI/NRAO and NAOJ. The National Radio Astronomy Observatory is a facility of the National Science Foundation operated under cooperative agreement by Associated Universities, Inc.






\vspace{5mm}
\facilities{ALMA}


\software{Astropy \citep{astropy_2013, astropy_2018},
          \texttt{CASA} \citep{McMullin_2007},
          \texttt{\textsuperscript{3D}BAROLO} \citep{3db_2015}
          }



\bibliography{aastex_references}{}

\begin{thebibliography}{}
\expandafter\ifx\csname natexlab\endcsname\relax\def\natexlab#1{#1}\fi
\providecommand{\url}[1]{\href{#1}{#1}}
\providecommand{\dodoi}[1]{doi:~\href{http://doi.org/#1}{\nolinkurl{#1}}}
\providecommand{\doeprint}[1]{\href{http://ascl.net/#1}{\nolinkurl{http://ascl.net/#1}}}
\providecommand{\doarXiv}[1]{\href{https://arxiv.org/abs/#1}{\nolinkurl{https://arxiv.org/abs/#1}}}

\bibitem[{{Astropy Collaboration} {et~al.}(2013){Astropy Collaboration}, {Robitaille}, {Tollerud}, {Greenfield}, {Droettboom}, {Bray}, {Aldcroft}, {Davis}, {Ginsburg}, {Price-Whelan}, {Kerzendorf}, {Conley}, {Crighton}, {Barbary}, {Muna}, {Ferguson}, {Grollier}, {Parikh}, {Nair}, {Unther}, {Deil}, {Woillez}, {Conseil}, {Kramer}, {Turner}, {Singer}, {Fox}, {Weaver}, {Zabalza}, {Edwards}, {Azalee Bostroem}, {Burke}, {Casey}, {Crawford}, {Dencheva}, {Ely}, {Jenness}, {Labrie}, {Lim}, {Pierfederici}, {Pontzen}, {Ptak}, {Refsdal}, {Servillat}, \& {Streicher}}]{astropy_2013}
{Astropy Collaboration}, {Robitaille}, T.~P., {Tollerud}, E.~J., {et~al.} 2013, \aap, 558, A33, \dodoi{10.1051/0004-6361/201322068}

\bibitem[{{Astropy Collaboration} {et~al.}(2018){Astropy Collaboration}, {Price-Whelan}, {Sip{\H{o}}cz}, {G{\"u}nther}, {Lim}, {Crawford}, {Conseil}, {Shupe}, {Craig}, {Dencheva}, {Ginsburg}, {VanderPlas}, {Bradley}, {P{\'e}rez-Su{\'a}rez}, {de Val-Borro}, {Aldcroft}, {Cruz}, {Robitaille}, {Tollerud}, {Ardelean}, {Babej}, {Bach}, {Bachetti}, {Bakanov}, {Bamford}, {Barentsen}, {Barmby}, {Baumbach}, {Berry}, {Biscani}, {Boquien}, {Bostroem}, {Bouma}, {Brammer}, {Bray}, {Breytenbach}, {Buddelmeijer}, {Burke}, {Calderone}, {Cano Rodr{\'\i}guez}, {Cara}, {Cardoso}, {Cheedella}, {Copin}, {Corrales}, {Crichton}, {D'Avella}, {Deil}, {Depagne}, {Dietrich}, {Donath}, {Droettboom}, {Earl}, {Erben}, {Fabbro}, {Ferreira}, {Finethy}, {Fox}, {Garrison}, {Gibbons}, {Goldstein}, {Gommers}, {Greco}, {Greenfield}, {Groener}, {Grollier}, {Hagen}, {Hirst}, {Homeier}, {Horton}, {Hosseinzadeh}, {Hu}, {Hunkeler}, {Ivezi{\'c}}, {Jain}, {Jenness}, {Kanarek}, {Kendrew}, {Kern}, {Kerzendorf}, {Khvalko}, {King}, {Kirkby}, {Kulkarni}, {Kumar}, {Lee}, {Lenz}, {Littlefair}, {Ma}, {Macleod}, {Mastropietro}, {McCully}, {Montagnac}, {Morris}, {Mueller}, {Mumford}, {Muna}, {Murphy}, {Nelson}, {Nguyen}, {Ninan}, {N{\"o}the}, {Ogaz}, {Oh}, {Parejko}, {Parley}, {Pascual}, {Patil}, {Patil}, {Plunkett}, {Prochaska}, {Rastogi}, {Reddy Janga}, {Sabater}, {Sakurikar}, {Seifert}, {Sherbert}, {Sherwood-Taylor}, {Shih}, {Sick}, {Silbiger}, {Singanamalla}, {Singer}, {Sladen}, {Sooley}, {Sornarajah}, {Streicher}, {Teuben}, {Thomas}, {Tremblay}, {Turner}, {Terr{\'o}n}, {van Kerkwijk}, {de la Vega}, {Watkins}, {Weaver}, {Whitmore}, {Woillez}, {Zabalza}, \& {Astropy Contributors}}]{astropy_2018}
{Astropy Collaboration}, {Price-Whelan}, A.~M., {Sip{\H{o}}cz}, B.~M., {et~al.} 2018, \aj, 156, 123, \dodoi{10.3847/1538-3881/aabc4f}

\bibitem[{{Begeman}(1989)}]{Begeman_1989}
{Begeman}, K.~G. 1989, \aap, 223, 47

\bibitem[{{B{\'e}thermin} {et~al.}(2020){B{\'e}thermin}, {Fudamoto}, {Ginolfi}, {Loiacono}, {Khusanova}, {Capak}, {Cassata}, {Faisst}, {Le F{\`e}vre}, {Schaerer}, {Silverman}, {Yan}, {Amorin}, {Bardelli}, {Boquien}, {Cimatti}, {Davidzon}, {Dessauges-Zavadsky}, {Fujimoto}, {Gruppioni}, {Hathi}, {Ibar}, {Jones}, {Koekemoer}, {Lagache}, {Lemaux}, {Moreau}, {Oesch}, {Pozzi}, {Riechers}, {Talia}, {Toft}, {Vallini}, {Vergani}, {Zamorani}, \& {Zucca}}]{Bethermin_2020}
{B{\'e}thermin}, M., {Fudamoto}, Y., {Ginolfi}, M., {et~al.} 2020, \aap, 643, A2, \dodoi{10.1051/0004-6361/202037649}

\bibitem[{{Burkert} {et~al.}(2010){Burkert}, {Genzel}, {Bouch{\'e}}, {Cresci}, {Khochfar}, {Sommer-Larsen}, {Sternberg}, {Naab}, {F{\"o}rster Schreiber}, {Tacconi}, {Shapiro}, {Hicks}, {Lutz}, {Davies}, {Buschkamp}, \& {Genel}}]{Burkert_2010}
{Burkert}, A., {Genzel}, R., {Bouch{\'e}}, N., {et~al.} 2010, \apj, 725, 2324, \dodoi{10.1088/0004-637X/725/2/2324}

\bibitem[{{Carilli} \& {Walter}(2013)}]{Carilli_2013}
{Carilli}, C.~L., \& {Walter}, F. 2013, \araa, 51, 105, \dodoi{10.1146/annurev-astro-082812-140953}

\bibitem[{{Carlstrom} {et~al.}(2011){Carlstrom}, {Ade}, {Aird}, {Benson}, {Bleem}, {Busetti}, {Chang}, {Chauvin}, {Cho}, {Crawford}, {Crites}, {Dobbs}, {Halverson}, {Heimsath}, {Holzapfel}, {Hrubes}, {Joy}, {Keisler}, {Lanting}, {Lee}, {Leitch}, {Leong}, {Lu}, {Lueker}, {Luong-Van}, {McMahon}, {Mehl}, {Meyer}, {Mohr}, {Montroy}, {Padin}, {Plagge}, {Pryke}, {Ruhl}, {Schaffer}, {Schwan}, {Shirokoff}, {Spieler}, {Staniszewski}, {Stark}, {Tucker}, {Vanderlinde}, {Vieira}, \& {Williamson}}]{Carlstrom_2011}
{Carlstrom}, J.~E., {Ade}, P.~A.~R., {Aird}, K.~A., {et~al.} 2011, \pasp, 123, 568, \dodoi{10.1086/659879}

\bibitem[{{Casey} {et~al.}(2014){Casey}, {Narayanan}, \& {Cooray}}]{Casey_2014}
{Casey}, C.~M., {Narayanan}, D., \& {Cooray}, A. 2014, \physrep, 541, 45, \dodoi{10.1016/j.physrep.2014.02.009}

\bibitem[{{Chiang} {et~al.}(2013){Chiang}, {Overzier}, \& {Gebhardt}}]{Chiang_2013}
{Chiang}, Y.-K., {Overzier}, R., \& {Gebhardt}, K. 2013, \apj, 779, 127, \dodoi{10.1088/0004-637X/779/2/127}

\bibitem[{{Dannerbauer} {et~al.}(2014){Dannerbauer}, {Kurk}, {De Breuck}, {Wylezalek}, {Santos}, {Koyama}, {Seymour}, {Tanaka}, {Hatch}, {Altieri}, {Coia}, {Galametz}, {Kodama}, {Miley}, {R{\"o}ttgering}, {Sanchez-Portal}, {Valtchanov}, {Venemans}, \& {Ziegler}}]{Dannenbauer_2014}
{Dannerbauer}, H., {Kurk}, J.~D., {De Breuck}, C., {et~al.} 2014, \aap, 570, A55, \dodoi{10.1051/0004-6361/201423771}

\bibitem[{{Davies} {et~al.}(2011){Davies}, {F{\"o}rster Schreiber}, {Cresci}, {Genzel}, {Bouch{\'e}}, {Burkert}, {Buschkamp}, {Genel}, {Hicks}, {Kurk}, {Lutz}, {Newman}, {Shapiro}, {Sternberg}, {Tacconi}, \& {Wuyts}}]{Davies_2011}
{Davies}, R., {F{\"o}rster Schreiber}, N.~M., {Cresci}, G., {et~al.} 2011, \apj, 741, 69, \dodoi{10.1088/0004-637X/741/2/69}

\bibitem[{{de Blok} {et~al.}(2008){de Blok}, {Walter}, {Brinks}, {Trachternach}, {Oh}, \& {Kennicutt}}]{deBlok_2008}
{de Blok}, W.~J.~G., {Walter}, F., {Brinks}, E., {et~al.} 2008, \aj, 136, 2648, \dodoi{10.1088/0004-6256/136/6/2648}

\bibitem[{{De Looze} {et~al.}(2011){De Looze}, {Baes}, {Bendo}, {Cortese}, \& {Fritz}}]{De_looze_2011}
{De Looze}, I., {Baes}, M., {Bendo}, G.~J., {Cortese}, L., \& {Fritz}, J. 2011, \mnras, 416, 2712, \dodoi{10.1111/j.1365-2966.2011.19223.x}

\bibitem[{{Dekel} \& {Birnboim}(2006)}]{Dekel_2006}
{Dekel}, A., \& {Birnboim}, Y. 2006, \mnras, 368, 2, \dodoi{10.1111/j.1365-2966.2006.10145.x}

\bibitem[{{Dekel} \& {Burkert}(2014)}]{Dekel_2014}
{Dekel}, A., \& {Burkert}, A. 2014, \mnras, 438, 1870, \dodoi{10.1093/mnras/stt2331}

\bibitem[{{Dekel} {et~al.}(2020){Dekel}, {Ginzburg}, {Jiang}, {Freundlich}, {Lapiner}, {Ceverino}, \& {Primack}}]{Dekel_2020a}
{Dekel}, A., {Ginzburg}, O., {Jiang}, F., {et~al.} 2020, \mnras, 493, 4126, \dodoi{10.1093/mnras/staa470}

\bibitem[{{Dekel} {et~al.}(2009{\natexlab{a}}){Dekel}, {Sari}, \& {Ceverino}}]{Dekel_2009a}
{Dekel}, A., {Sari}, R., \& {Ceverino}, D. 2009{\natexlab{a}}, \apj, 703, 785, \dodoi{10.1088/0004-637X/703/1/785}

\bibitem[{{Dekel} {et~al.}(2009{\natexlab{b}}){Dekel}, {Birnboim}, {Engel}, {Freundlich}, {Goerdt}, {Mumcuoglu}, {Neistein}, {Pichon}, {Teyssier}, \& {Zinger}}]{Dekel_2009b}
{Dekel}, A., {Birnboim}, Y., {Engel}, G., {et~al.} 2009{\natexlab{b}}, \nat, 457, 451, \dodoi{10.1038/nature07648}

\bibitem[{Di~Teodoro(2015)}]{3db_online}
Di~Teodoro, E. 2015, 3D-BAROLO Documentation.
\newblock \url{https://bbarolo.readthedocs.io/en/latest/}

\bibitem[{{Di Teodoro} \& {Fraternali}(2015)}]{3db_2015}
{Di Teodoro}, E.~M., \& {Fraternali}, F. 2015, \mnras, 451, 3021, \dodoi{10.1093/mnras/stv1213}

\bibitem[{{Di Teodoro} {et~al.}(2016){Di Teodoro}, {Fraternali}, \& {Miller}}]{Teodoro_2016}
{Di Teodoro}, E.~M., {Fraternali}, F., \& {Miller}, S.~H. 2016, \aap, 594, A77, \dodoi{10.1051/0004-6361/201628315}

\bibitem[{{Eales} {et~al.}(2010){Eales}, {Dunne}, {Clements}, {Cooray}, {De Zotti}, {Dye}, {Ivison}, {Jarvis}, {Lagache}, {Maddox}, {Negrello}, {Serjeant}, {Thompson}, {Van Kampen}, {Amblard}, {Andreani}, {Baes}, {Beelen}, {Bendo}, {Benford}, {Bertoldi}, {Bock}, {Bonfield}, {Boselli}, {Bridge}, {Buat}, {Burgarella}, {Carlberg}, {Cava}, {Chanial}, {Charlot}, {Christopher}, {Coles}, {Cortese}, {Dariush}, {da Cunha}, {Dalton}, {Danese}, {Dannerbauer}, {Driver}, {Dunlop}, {Fan}, {Farrah}, {Frayer}, {Frenk}, {Geach}, {Gardner}, {Gomez}, {Gonz{\'a}lez-Nuevo}, {Gonz{\'a}lez-Solares}, {Griffin}, {Hardcastle}, {Hatziminaoglou}, {Herranz}, {Hughes}, {Ibar}, {Jeong}, {Lacey}, {Lapi}, {Lawrence}, {Lee}, {Leeuw}, {Liske}, {L{\'o}pez-Caniego}, {M{\"u}ller}, {Nandra}, {Panuzzo}, {Papageorgiou}, {Patanchon}, {Peacock}, {Pearson}, {Phillipps}, {Pohlen}, {Popescu}, {Rawlings}, {Rigby}, {Rigopoulou}, {Robotham}, {Rodighiero}, {Sansom}, {Schulz}, {Scott}, {Smith}, {Sibthorpe}, {Smail}, {Stevens}, {Sutherland}, {Takeuchi}, {Tedds}, {Temi}, {Tuffs}, {Trichas}, {Vaccari}, {Valtchanov}, {van der Werf}, {Verma}, {Vieria}, {Vlahakis}, \& {White}}]{Eales_2010}
{Eales}, S., {Dunne}, L., {Clements}, D., {et~al.} 2010, \pasp, 122, 499, \dodoi{10.1086/653086}

\bibitem[{{Fall} \& {Efstathiou}(1980)}]{Fall_1980}
{Fall}, S.~M., \& {Efstathiou}, G. 1980, \mnras, 193, 189, \dodoi{10.1093/mnras/193.2.189}

\bibitem[{{F{\"o}rster Schreiber} \& {Wuyts}(2020)}]{Foerster_schreiber_2020}
{F{\"o}rster Schreiber}, N.~M., \& {Wuyts}, S. 2020, \araa, 58, 661, \dodoi{10.1146/annurev-astro-032620-021910}

\bibitem[{{F{\"o}rster Schreiber} {et~al.}(2009){F{\"o}rster Schreiber}, {Genzel}, {Bouch{\'e}}, {Cresci}, {Davies}, {Buschkamp}, {Shapiro}, {Tacconi}, {Hicks}, {Genel}, {Shapley}, {Erb}, {Steidel}, {Lutz}, {Eisenhauer}, {Gillessen}, {Sternberg}, {Renzini}, {Cimatti}, {Daddi}, {Kurk}, {Lilly}, {Kong}, {Lehnert}, {Nesvadba}, {Verma}, {McCracken}, {Arimoto}, {Mignoli}, \& {Onodera}}]{Foerster_Schreiber_2009}
{F{\"o}rster Schreiber}, N.~M., {Genzel}, R., {Bouch{\'e}}, N., {et~al.} 2009, \apj, 706, 1364, \dodoi{10.1088/0004-637X/706/2/1364}

\bibitem[{{F{\"o}rster Schreiber} {et~al.}(2018){F{\"o}rster Schreiber}, {Renzini}, {Mancini}, {Genzel}, {Bouch{\'e}}, {Cresci}, {Hicks}, {Lilly}, {Peng}, {Burkert}, {Carollo}, {Cimatti}, {Daddi}, {Davies}, {Genel}, {Kurk}, {Lang}, {Lutz}, {Mainieri}, {McCracken}, {Mignoli}, {Naab}, {Oesch}, {Pozzetti}, {Scodeggio}, {Shapiro Griffin}, {Shapley}, {Sternberg}, {Tacchella}, {Tacconi}, {Wuyts}, \& {Zamorani}}]{Foerster_Schreiber_2018}
{F{\"o}rster Schreiber}, N.~M., {Renzini}, A., {Mancini}, C., {et~al.} 2018, \apjs, 238, 21, \dodoi{10.3847/1538-4365/aadd49}

\bibitem[{{Franx} {et~al.}(1991){Franx}, {Illingworth}, \& {de Zeeuw}}]{Franx_1991}
{Franx}, M., {Illingworth}, G., \& {de Zeeuw}, T. 1991, \apj, 383, 112, \dodoi{10.1086/170769}

\bibitem[{{Fraternali} {et~al.}(2021){Fraternali}, {Karim}, {Magnelli}, {G{\'o}mez-Guijarro}, {Jim{\'e}nez-Andrade}, \& {Posses}}]{Fratenalli_2021}
{Fraternali}, F., {Karim}, A., {Magnelli}, B., {et~al.} 2021, \aap, 647, A194, \dodoi{10.1051/0004-6361/202039807}

\bibitem[{{Ginolfi} {et~al.}(2020){Ginolfi}, {Jones}, {B{\'e}thermin}, {Fudamoto}, {Loiacono}, {Fujimoto}, {Le F{\'e}vre}, {Faisst}, {Schaerer}, {Cassata}, {Silverman}, {Yan}, {Capak}, {Bardelli}, {Boquien}, {Carraro}, {Dessauges-Zavadsky}, {Giavalisco}, {Gruppioni}, {Ibar}, {Khusanova}, {Lemaux}, {Maiolino}, {Narayanan}, {Oesch}, {Pozzi}, {Rodighiero}, {Talia}, {Toft}, {Vallini}, {Vergani}, \& {Zamorani}}]{Ginolfi_2020}
{Ginolfi}, M., {Jones}, G.~C., {B{\'e}thermin}, M., {et~al.} 2020, \aap, 633, A90, \dodoi{10.1051/0004-6361/201936872}

\bibitem[{{Grand} {et~al.}(2017){Grand}, {G{\'o}mez}, {Marinacci}, {Pakmor}, {Springel}, {Campbell}, {Frenk}, {Jenkins}, \& {White}}]{Grand_2017}
{Grand}, R. J.~J., {G{\'o}mez}, F.~A., {Marinacci}, F., {et~al.} 2017, \mnras, 467, 179, \dodoi{10.1093/mnras/stx071}

\bibitem[{{Gullberg} {et~al.}(2019){Gullberg}, {Smail}, {Swinbank}, {Dudzevi{\v{c}}i{\={u}}t{\.{e}}}, {Stach}, {Thomson}, {Almaini}, {Chen}, {Conselice}, {Cooke}, {Farrah}, {Ivison}, {Maltby}, {Micha{\l}owski}, {Simpson}, {Scott}, {Wardlow}, \& {Weiss}}]{Gullberg_2019}
{Gullberg}, B., {Smail}, I., {Swinbank}, A.~M., {et~al.} 2019, \mnras, 490, 4956, \dodoi{10.1093/mnras/stz2835}

\bibitem[{{Hayward} \& {Hopkins}(2017)}]{Hayward_2017}
{Hayward}, C.~C., \& {Hopkins}, P.~F. 2017, \mnras, 465, 1682, \dodoi{10.1093/mnras/stw2888}

\bibitem[{{Herrera-Camus} {et~al.}(2015){Herrera-Camus}, {Bolatto}, {Wolfire}, {Smith}, {Croxall}, {Kennicutt}, {Calzetti}, {Helou}, {Walter}, {Leroy}, {Draine}, {Brandl}, {Armus}, {Sandstrom}, {Dale}, {Aniano}, {Meidt}, {Boquien}, {Hunt}, {Galametz}, {Tabatabaei}, {Murphy}, {Appleton}, {Roussel}, {Engelbracht}, \& {Beirao}}]{Herrera_Camus_2015}
{Herrera-Camus}, R., {Bolatto}, A.~D., {Wolfire}, M.~G., {et~al.} 2015, \apj, 800, 1, \dodoi{10.1088/0004-637X/800/1/1}

\bibitem[{{Hill} {et~al.}(2020){Hill}, {Chapman}, {Scott}, {Apostolovski}, {Aravena}, {B{\'e}thermin}, {Bradford}, {Canning}, {De Breuck}, {Dong}, {Gonzalez}, {Greve}, {Hayward}, {Hezaveh}, {Litke}, {Malkan}, {Marrone}, {Phadke}, {Reuter}, {Rotermund}, {Spilker}, {Vieira}, \& {Wei{\ss}}}]{Hill_2020}
{Hill}, R., {Chapman}, S., {Scott}, D., {et~al.} 2020, \mnras, 495, 3124, \dodoi{10.1093/mnras/staa1275}

\bibitem[{{Hill} {et~al.}(2022){Hill}, {Chapman}, {Phadke}, {Aravena}, {Archipley}, {Ashby}, {B{\'e}thermin}, {Canning}, {Gonzalez}, {Greve}, {Gururajan}, {Hayward}, {Hezaveh}, {Jarugula}, {MacIntyre}, {Marrone}, {Miller}, {Rennehan}, {Reuter}, {Rotermund}, {Scott}, {Spilker}, {Vieira}, {Wang}, \& {Wei{\ss}}}]{Hill_2022}
{Hill}, R., {Chapman}, S., {Phadke}, K.~A., {et~al.} 2022, \mnras, 512, 4352, \dodoi{10.1093/mnras/stab3539}

\bibitem[{{Hodge} {et~al.}(2012){Hodge}, {Carilli}, {Walter}, {de Blok}, {Riechers}, {Daddi}, \& {Lentati}}]{Hodge_2012}
{Hodge}, J.~A., {Carilli}, C.~L., {Walter}, F., {et~al.} 2012, \apj, 760, 11, \dodoi{10.1088/0004-637X/760/1/11}

\bibitem[{{Jones} {et~al.}(2021){Jones}, {Vergani}, {Romano}, {Ginolfi}, {Fudamoto}, {B{\'e}thermin}, {Fujimoto}, {Lemaux}, {Morselli}, {Capak}, {Cassata}, {Faisst}, {Le F{\`e}vre}, {Schaerer}, {Silverman}, {Yan}, {Boquien}, {Cimatti}, {Dessauges-Zavadsky}, {Ibar}, {Maiolino}, {Rizzo}, {Talia}, \& {Zamorani}}]{Jones_2021}
{Jones}, G.~C., {Vergani}, D., {Romano}, M., {et~al.} 2021, \mnras, 507, 3540, \dodoi{10.1093/mnras/stab2226}

\bibitem[{{J{\'o}zsa} {et~al.}(2007){J{\'o}zsa}, {Kenn}, {Klein}, \& {Oosterloo}}]{Jozsa_2007}
{J{\'o}zsa}, G.~I.~G., {Kenn}, F., {Klein}, U., \& {Oosterloo}, T.~A. 2007, \aap, 468, 731, \dodoi{10.1051/0004-6361:20066164}

\bibitem[{{Kere{\v{s}}} {et~al.}(2005){Kere{\v{s}}}, {Katz}, {Weinberg}, \& {Dav{\'e}}}]{Keres_2005}
{Kere{\v{s}}}, D., {Katz}, N., {Weinberg}, D.~H., \& {Dav{\'e}}, R. 2005, \mnras, 363, 2, \dodoi{10.1111/j.1365-2966.2005.09451.x}

\bibitem[{{King} \& {Pounds}(2015)}]{King_2015}
{King}, A., \& {Pounds}, K. 2015, \araa, 53, 115, \dodoi{10.1146/annurev-astro-082214-122316}

\bibitem[{{Kohandel} {et~al.}(2019){Kohandel}, {Pallottini}, {Ferrara}, {Zanella}, {Behrens}, {Carniani}, {Gallerani}, \& {Vallini}}]{Kohandel_2019}
{Kohandel}, M., {Pallottini}, A., {Ferrara}, A., {et~al.} 2019, \mnras, 487, 3007, \dodoi{10.1093/mnras/stz1486}

\bibitem[{{Kohandel} {et~al.}(2023){Kohandel}, {Pallottini}, {Ferrara}, {Zanella}, {Rizzo}, \& {Carniani}}]{Kohandel_2023}
---. 2023, arXiv e-prints, arXiv:2311.05832, \dodoi{10.48550/arXiv.2311.05832}

\bibitem[{{Krajnovi{\'c}} {et~al.}(2006){Krajnovi{\'c}}, {Cappellari}, {de Zeeuw}, \& {Copin}}]{Krajnovic_2006}
{Krajnovi{\'c}}, D., {Cappellari}, M., {de Zeeuw}, P.~T., \& {Copin}, Y. 2006, \mnras, 366, 787, \dodoi{10.1111/j.1365-2966.2005.09902.x}

\bibitem[{{Kretschmer} {et~al.}(2022){Kretschmer}, {Dekel}, \& {Teyssier}}]{Kretschmer_2022}
{Kretschmer}, M., {Dekel}, A., \& {Teyssier}, R. 2022, \mnras, 510, 3266, \dodoi{10.1093/mnras/stab3648}

\bibitem[{{Krumholz} {et~al.}(2018){Krumholz}, {Burkhart}, {Forbes}, \& {Crocker}}]{Krumholz_2018}
{Krumholz}, M.~R., {Burkhart}, B., {Forbes}, J.~C., \& {Crocker}, R.~M. 2018, \mnras, 477, 2716, \dodoi{10.1093/mnras/sty852}

\bibitem[{{Lagache} {et~al.}(2018){Lagache}, {Cousin}, \& {Chatzikos}}]{Lagache_2018}
{Lagache}, G., {Cousin}, M., \& {Chatzikos}, M. 2018, \aap, 609, A130, \dodoi{10.1051/0004-6361/201732019}

\bibitem[{{Lelli} {et~al.}(2018){Lelli}, {De Breuck}, {Falkendal}, {Fraternali}, {Man}, {Nesvadba}, \& {Lehnert}}]{Lelli_2018}
{Lelli}, F., {De Breuck}, C., {Falkendal}, T., {et~al.} 2018, \mnras, 479, 5440, \dodoi{10.1093/mnras/sty1795}

\bibitem[{{Lelli} {et~al.}(2021){Lelli}, {Di Teodoro}, {Fraternali}, {Man}, {Zhang}, {De Breuck}, {Davis}, \& {Maiolino}}]{Lelli_2021}
{Lelli}, F., {Di Teodoro}, E.~M., {Fraternali}, F., {et~al.} 2021, Science, 371, 713, \dodoi{10.1126/science.abc1893}

\bibitem[{{Lelli} {et~al.}(2016){Lelli}, {McGaugh}, \& {Schombert}}]{Lelli_2016}
{Lelli}, F., {McGaugh}, S.~S., \& {Schombert}, J.~M. 2016, \aj, 152, 157, \dodoi{10.3847/0004-6256/152/6/157}

\bibitem[{{Litke} {et~al.}(2019){Litke}, {Marrone}, {Spilker}, {Aravena}, {B{\'e}thermin}, {Chapman}, {Chen}, {de Breuck}, {Dong}, {Gonzalez}, {Greve}, {Hayward}, {Hezaveh}, {Jarugula}, {Ma}, {Morningstar}, {Narayanan}, {Phadke}, {Reuter}, {Vieira}, \& {Weiss}}]{Litke_2019}
{Litke}, K.~C., {Marrone}, D.~P., {Spilker}, J.~S., {et~al.} 2019, \apj, 870, 80, \dodoi{10.3847/1538-4357/aaf057}

\bibitem[{{Madau} \& {Dickinson}(2014)}]{Madau_2014}
{Madau}, P., \& {Dickinson}, M. 2014, \araa, 52, 415, \dodoi{10.1146/annurev-astro-081811-125615}

\bibitem[{{Mart{\'\i}nez-Garc{\'\i}a} {et~al.}(2023){Mart{\'\i}nez-Garc{\'\i}a}, {del Pino}, \& {Aparicio}}]{MartinezGarcia_2023}
{Mart{\'\i}nez-Garc{\'\i}a}, A.~M., {del Pino}, A., \& {Aparicio}, A. 2023, \mnras, 518, 3083, \dodoi{10.1093/mnras/stac3305}

\bibitem[{{McMullin} {et~al.}(2007){McMullin}, {Waters}, {Schiebel}, {Young}, \& {Golap}}]{McMullin_2007}
{McMullin}, J.~P., {Waters}, B., {Schiebel}, D., {Young}, W., \& {Golap}, K. 2007, in Astronomical Society of the Pacific Conference Series, Vol. 376, Astronomical Data Analysis Software and Systems XVI, ed. R.~A. {Shaw}, F.~{Hill}, \& D.~J. {Bell}, 127

\bibitem[{{Miley} \& {De Breuck}(2008)}]{Miley_2008}
{Miley}, G., \& {De Breuck}, C. 2008, \aapr, 15, 67, \dodoi{10.1007/s00159-007-0008-z}

\bibitem[{{Miller} {et~al.}(2018){Miller}, {Chapman}, {Aravena}, {Ashby}, {Hayward}, {Vieira}, {Wei{\ss}}, {Babul}, {B{\'e}thermin}, {Bradford}, {Brodwin}, {Carlstrom}, {Chen}, {Cunningham}, {De Breuck}, {Gonzalez}, {Greve}, {Harnett}, {Hezaveh}, {Lacaille}, {Litke}, {Ma}, {Malkan}, {Marrone}, {Morningstar}, {Murphy}, {Narayanan}, {Pass}, {Perry}, {Phadke}, {Rennehan}, {Rotermund}, {Simpson}, {Spilker}, {Sreevani}, {Stark}, {Strandet}, \& {Strom}}]{Miller_2018}
{Miller}, T.~B., {Chapman}, S.~C., {Aravena}, M., {et~al.} 2018, \nat, 556, 469, \dodoi{10.1038/s41586-018-0025-2}

\bibitem[{{Mocanu} {et~al.}(2013){Mocanu}, {Crawford}, {Vieira}, {Aird}, {Aravena}, {Austermann}, {Benson}, {B{\'e}thermin}, {Bleem}, {Bothwell}, {Carlstrom}, {Chang}, {Chapman}, {Cho}, {Crites}, {de Haan}, {Dobbs}, {Everett}, {George}, {Halverson}, {Harrington}, {Hezaveh}, {Holder}, {Holzapfel}, {Hoover}, {Hrubes}, {Keisler}, {Knox}, {Lee}, {Leitch}, {Lueker}, {Luong-Van}, {Marrone}, {McMahon}, {Mehl}, {Meyer}, {Mohr}, {Montroy}, {Natoli}, {Padin}, {Plagge}, {Pryke}, {Rest}, {Reichardt}, {Ruhl}, {Sayre}, {Schaffer}, {Shirokoff}, {Spieler}, {Spilker}, {Stalder}, {Staniszewski}, {Stark}, {Story}, {Switzer}, {Vanderlinde}, \& {Williamson}}]{Mocanu_2013}
{Mocanu}, L.~M., {Crawford}, T.~M., {Vieira}, J.~D., {et~al.} 2013, \apj, 779, 61, \dodoi{10.1088/0004-637X/779/1/61}

\bibitem[{Murdin(2001)}]{Murdin_2001}
Murdin, P. 2001, Encyclopedia of Astronomy \& Astrophysics (CRC Press).
\newblock \url{https://books.google.co.in/books?id=R5dBEAAAQBAJ}

\bibitem[{{Neeleman} {et~al.}(2020){Neeleman}, {Prochaska}, {Kanekar}, \& {Rafelski}}]{Neeleman_2020}
{Neeleman}, M., {Prochaska}, J.~X., {Kanekar}, N., \& {Rafelski}, M. 2020, \nat, 581, 269, \dodoi{10.1038/s41586-020-2276-y}

\bibitem[{{Neri} {et~al.}(2014){Neri}, {Downes}, {Cox}, \& {Walter}}]{Neri_2014}
{Neri}, R., {Downes}, D., {Cox}, P., \& {Walter}, F. 2014, \aap, 562, A35, \dodoi{10.1051/0004-6361/201322528}

\bibitem[{{Oteo} {et~al.}(2016){Oteo}, {Ivison}, {Dunne}, {Smail}, {Swinbank}, {Zhang}, {Lewis}, {Maddox}, {Riechers}, {Serjeant}, {Van der Werf}, {Biggs}, {Bremer}, {Cigan}, {Clements}, {Cooray}, {Dannerbauer}, {Eales}, {Ibar}, {Messias}, {Micha{\l}owski}, {P{\'e}rez-Fournon}, \& {van Kampen}}]{Oteo_2016}
{Oteo}, I., {Ivison}, R.~J., {Dunne}, L., {et~al.} 2016, \apj, 827, 34, \dodoi{10.3847/0004-637X/827/1/34}

\bibitem[{{Oteo} {et~al.}(2018){Oteo}, {Ivison}, {Dunne}, {Manilla-Robles}, {Maddox}, {Lewis}, {de Zotti}, {Bremer}, {Clements}, {Cooray}, {Dannerbauer}, {Eales}, {Greenslade}, {Omont}, {Perez{\textendash}Fourn{\'o}n}, {Riechers}, {Scott}, {van der Werf}, {Weiss}, \& {Zhang}}]{Oteo_2018}
---. 2018, \apj, 856, 72, \dodoi{10.3847/1538-4357/aaa1f1}

\bibitem[{{Overzier}(2016)}]{Overzier_2016}
{Overzier}, R.~A. 2016, \aapr, 24, 14, \dodoi{10.1007/s00159-016-0100-3}

\bibitem[{{Pallottini} {et~al.}(2022){Pallottini}, {Ferrara}, {Gallerani}, {Behrens}, {Kohandel}, {Carniani}, {Vallini}, {Salvadori}, {Gelli}, {Sommovigo}, {D'Odorico}, {Di Mascia}, \& {Pizzati}}]{Pallottini_2022}
{Pallottini}, A., {Ferrara}, A., {Gallerani}, S., {et~al.} 2022, \mnras, 513, 5621, \dodoi{10.1093/mnras/stac1281}

\bibitem[{{Pilbratt} {et~al.}(2010){Pilbratt}, {Riedinger}, {Passvogel}, {Crone}, {Doyle}, {Gageur}, {Heras}, {Jewell}, {Metcalfe}, {Ott}, \& {Schmidt}}]{Herschel_2010}
{Pilbratt}, G.~L., {Riedinger}, J.~R., {Passvogel}, T., {et~al.} 2010, \aap, 518, L1, \dodoi{10.1051/0004-6361/201014759}

\bibitem[{{Pillepich} {et~al.}(2019){Pillepich}, {Nelson}, {Springel}, {Pakmor}, {Torrey}, {Weinberger}, {Vogelsberger}, {Marinacci}, {Genel}, {van der Wel}, \& {Hernquist}}]{Pillepich_2019}
{Pillepich}, A., {Nelson}, D., {Springel}, V., {et~al.} 2019, \mnras, 490, 3196, \dodoi{10.1093/mnras/stz2338}

\bibitem[{{Planck Collaboration} {et~al.}(2016){Planck Collaboration}, {Ade}, {Aghanim}, {Arnaud}, {Ashdown}, {Aumont}, {Baccigalupi}, {Banday}, {Barreiro}, {Bartlett}, {Bartolo}, {Battaner}, {Battye}, {Benabed}, {Beno{\^\i}t}, {Benoit-L{\'e}vy}, {Bernard}, {Bersanelli}, {Bielewicz}, {Bock}, {Bonaldi}, {Bonavera}, {Bond}, {Borrill}, {Bouchet}, {Boulanger}, {Bucher}, {Burigana}, {Butler}, {Calabrese}, {Cardoso}, {Catalano}, {Challinor}, {Chamballu}, {Chary}, {Chiang}, {Chluba}, {Christensen}, {Church}, {Clements}, {Colombi}, {Colombo}, {Combet}, {Coulais}, {Crill}, {Curto}, {Cuttaia}, {Danese}, {Davies}, {Davis}, {de Bernardis}, {de Rosa}, {de Zotti}, {Delabrouille}, {D{\'e}sert}, {Di Valentino}, {Dickinson}, {Diego}, {Dolag}, {Dole}, {Donzelli}, {Dor{\'e}}, {Douspis}, {Ducout}, {Dunkley}, {Dupac}, {Efstathiou}, {Elsner}, {En{\ss}lin}, {Eriksen}, {Farhang}, {Fergusson}, {Finelli}, {Forni}, {Frailis}, {Fraisse}, {Franceschi}, {Frejsel}, {Galeotta}, {Galli}, {Ganga}, {Gauthier}, {Gerbino}, {Ghosh}, {Giard}, {Giraud-H{\'e}raud}, {Giusarma}, {Gjerl{\o}w}, {Gonz{\'a}lez-Nuevo}, {G{\'o}rski}, {Gratton}, {Gregorio}, {Gruppuso}, {Gudmundsson}, {Hamann}, {Hansen}, {Hanson}, {Harrison}, {Helou}, {Henrot-Versill{\'e}}, {Hern{\'a}ndez-Monteagudo}, {Herranz}, {Hildebrandt}, {Hivon}, {Hobson}, {Holmes}, {Hornstrup}, {Hovest}, {Huang}, {Huffenberger}, {Hurier}, {Jaffe}, {Jaffe}, {Jones}, {Juvela}, {Keih{\"a}nen}, {Keskitalo}, {Kisner}, {Kneissl}, {Knoche}, {Knox}, {Kunz}, {Kurki-Suonio}, {Lagache}, {L{\"a}hteenm{\"a}ki}, {Lamarre}, {Lasenby}, {Lattanzi}, {Lawrence}, {Leahy}, {Leonardi}, {Lesgourgues}, {Levrier}, {Lewis}, {Liguori}, {Lilje}, {Linden-V{\o}rnle}, {L{\'o}pez-Caniego}, {Lubin}, {Mac{\'\i}as-P{\'e}rez}, {Maggio}, {Maino}, {Mandolesi}, {Mangilli}, {Marchini}, {Maris}, {Martin}, {Martinelli}, {Mart{\'\i}nez-Gonz{\'a}lez}, {Masi}, {Matarrese}, {McGehee}, {Meinhold}, {Melchiorri}, {Melin}, {Mendes}, {Mennella}, {Migliaccio}, {Millea}, {Mitra}, {Miville-Desch{\^e}nes}, {Moneti}, {Montier}, {Morgante}, {Mortlock}, {Moss}, {Munshi}, {Murphy}, {Naselsky}, {Nati}, {Natoli}, {Netterfield}, {N{\o}rgaard-Nielsen}, {Noviello}, {Novikov}, {Novikov}, {Oxborrow}, {Paci}, {Pagano}, {Pajot}, {Paladini}, {Paoletti}, {Partridge}, {Pasian}, {Patanchon}, {Pearson}, {Perdereau}, {Perotto}, {Perrotta}, {Pettorino}, {Piacentini}, {Piat}, {Pierpaoli}, {Pietrobon}, {Plaszczynski}, {Pointecouteau}, {Polenta}, {Popa}, {Pratt}, {Pr{\'e}zeau}, {Prunet}, {Puget}, {Rachen}, {Reach}, {Rebolo}, {Reinecke}, {Remazeilles}, {Renault}, {Renzi}, {Ristorcelli}, {Rocha}, {Rosset}, {Rossetti}, {Roudier}, {Rouill{\'e} d'Orfeuil}, {Rowan-Robinson}, {Rubi{\~n}o-Mart{\'\i}n}, {Rusholme}, {Said}, {Salvatelli}, {Salvati}, {Sandri}, {Santos}, {Savelainen}, {Savini}, {Scott}, {Seiffert}, {Serra}, {Shellard}, {Spencer}, {Spinelli}, {Stolyarov}, {Stompor}, {Sudiwala}, {Sunyaev}, {Sutton}, {Suur-Uski}, {Sygnet}, {Tauber}, {Terenzi}, {Toffolatti}, {Tomasi}, {Tristram}, {Trombetti}, {Tucci}, {Tuovinen}, {T{\"u}rler}, {Umana}, {Valenziano}, {Valiviita}, {Van Tent}, {Vielva}, {Villa}, {Wade}, {Wandelt}, {Wehus}, {White}, {White}, {Wilkinson}, {Yvon}, {Zacchei}, \& {Zonca}}]{Planck_2016}
{Planck Collaboration}, {Ade}, P.~A.~R., {Aghanim}, N., {et~al.} 2016, \aap, 594, A13, \dodoi{10.1051/0004-6361/201525830}

\bibitem[{{Posses} {et~al.}(2023){Posses}, {Aravena}, {Gonz{\'a}lez-L{\'o}pez}, {Assef}, {Lambert}, {Jones}, {Bouwens}, {Brisbin}, {D{\'\i}az-Santos}, {Herrera-Camus}, {Ricci}, \& {Smit}}]{Posses_2023}
{Posses}, A.~C., {Aravena}, M., {Gonz{\'a}lez-L{\'o}pez}, J., {et~al.} 2023, \aap, 669, A46, \dodoi{10.1051/0004-6361/202243399}

\bibitem[{{Rees} \& {Ostriker}(1977)}]{Rees_1977}
{Rees}, M.~J., \& {Ostriker}, J.~P. 1977, \mnras, 179, 541, \dodoi{10.1093/mnras/179.4.541}

\bibitem[{{Remus} {et~al.}(2022){Remus}, {Dolag}, \& {Dannerbauer}}]{Remus_2022}
{Remus}, R.-S., {Dolag}, K., \& {Dannerbauer}, H. 2022, arXiv e-prints, arXiv:2208.01053.
\newblock \doarXiv{2208.01053}

\bibitem[{{Rennehan} {et~al.}(2020){Rennehan}, {Babul}, {Hayward}, {Bottrell}, {Hani}, \& {Chapman}}]{Rennehan_2020}
{Rennehan}, D., {Babul}, A., {Hayward}, C.~C., {et~al.} 2020, \mnras, 493, 4607, \dodoi{10.1093/mnras/staa541}

\bibitem[{{Rizzo} {et~al.}(2022){Rizzo}, {Kohandel}, {Pallottini}, {Zanella}, {Ferrara}, {Vallini}, \& {Toft}}]{Rizzo_2022}
{Rizzo}, F., {Kohandel}, M., {Pallottini}, A., {et~al.} 2022, \aap, 667, A5, \dodoi{10.1051/0004-6361/202243582}

\bibitem[{{Rizzo} {et~al.}(2021){Rizzo}, {Vegetti}, {Fraternali}, {Stacey}, \& {Powell}}]{Rizzo_2021}
{Rizzo}, F., {Vegetti}, S., {Fraternali}, F., {Stacey}, H.~R., \& {Powell}, D. 2021, \mnras, 507, 3952, \dodoi{10.1093/mnras/stab2295}

\bibitem[{{Rizzo} {et~al.}(2020){Rizzo}, {Vegetti}, {Powell}, {Fraternali}, {McKean}, {Stacey}, \& {White}}]{Rizzo_2020}
{Rizzo}, F., {Vegetti}, S., {Powell}, D., {et~al.} 2020, \nat, 584, 201, \dodoi{10.1038/s41586-020-2572-6}

\bibitem[{{Rogstad} {et~al.}(1974){Rogstad}, {Lockhart}, \& {Wright}}]{Rogstad_1974}
{Rogstad}, D.~H., {Lockhart}, I.~A., \& {Wright}, M.~C.~H. 1974, \apj, 193, 309, \dodoi{10.1086/153164}

\bibitem[{{Roman-Oliveira} {et~al.}(2023){Roman-Oliveira}, {Fraternali}, \& {Rizzo}}]{Roman_Oliviera_2023}
{Roman-Oliveira}, F., {Fraternali}, F., \& {Rizzo}, F. 2023, \mnras, 521, 1045, \dodoi{10.1093/mnras/stad530}

\bibitem[{{Romanowsky} \& {Fall}(2012)}]{Romanowsky_2012}
{Romanowsky}, A.~J., \& {Fall}, S.~M. 2012, \apjs, 203, 17, \dodoi{10.1088/0067-0049/203/2/17}

\bibitem[{{Rotermund} {et~al.}(2021){Rotermund}, {Chapman}, {Phadke}, {Hill}, {Pass}, {Aravena}, {Ashby}, {Babul}, {B{\'e}thermin}, {Canning}, {de Breuck}, {Dong}, {Gonzalez}, {Hayward}, {Jarugula}, {Marrone}, {Narayanan}, {Reuter}, {Scott}, {Spilker}, {Vieira}, {Wang}, \& {Weiss}}]{Rotermund_2021}
{Rotermund}, K.~M., {Chapman}, S.~C., {Phadke}, K.~A., {et~al.} 2021, \mnras, 502, 1797, \dodoi{10.1093/mnras/stab103}

\bibitem[{{Sellwood} \& {Spekkens}(2015)}]{Sellwood_2015}
{Sellwood}, J.~A., \& {Spekkens}, K. 2015, arXiv e-prints, arXiv:1509.07120.
\newblock \doarXiv{1509.07120}

\bibitem[{{Sharda} {et~al.}(2019){Sharda}, {da Cunha}, {Federrath}, {Wisnioski}, {Di Teodoro}, {Tadaki}, {Yun}, {Aretxaga}, \& {Kawabe}}]{Sharda_2019}
{Sharda}, P., {da Cunha}, E., {Federrath}, C., {et~al.} 2019, \mnras, 487, 4305, \dodoi{10.1093/mnras/stz1543}

\bibitem[{Sicking(1997)}]{Sicking_1997}
Sicking, F. 1997, PhD thesis, University of Groningen

\bibitem[{{Solomon} {et~al.}(1997){Solomon}, {Downes}, {Radford}, \& {Barrett}}]{Solomon_1997}
{Solomon}, P.~M., {Downes}, D., {Radford}, S.~J.~E., \& {Barrett}, J.~W. 1997, \apj, 478, 144, \dodoi{10.1086/303765}

\bibitem[{{Swaters} {et~al.}(2009){Swaters}, {Sancisi}, {van Albada}, \& {van der Hulst}}]{Swaters_2009}
{Swaters}, R.~A., {Sancisi}, R., {van Albada}, T.~S., \& {van der Hulst}, J.~M. 2009, \aap, 493, 871, \dodoi{10.1051/0004-6361:200810516}

\bibitem[{{Swinbank} {et~al.}(2009){Swinbank}, {Webb}, {Richard}, {Bower}, {Ellis}, {Illingworth}, {Jones}, {Kriek}, {Smail}, {Stark}, \& {van Dokkum}}]{Swinbank_2009}
{Swinbank}, A.~M., {Webb}, T.~M., {Richard}, J., {et~al.} 2009, \mnras, 400, 1121, \dodoi{10.1111/j.1365-2966.2009.15617.x}

\bibitem[{{Tadaki} {et~al.}(2019){Tadaki}, {Kodama}, {Hayashi}, {Shimakawa}, {Koyama}, {Lee}, {Tanaka}, {Hatsukade}, {Iono}, {Kohno}, {Matsuda}, {Suzuki}, {Tamura}, {Toshikawa}, \& {Umehata}}]{Tadaki_2019}
{Tadaki}, K.-i., {Kodama}, T., {Hayashi}, M., {et~al.} 2019, \pasj, 71, 40, \dodoi{10.1093/pasj/psz005}

\bibitem[{{Toomre}(1964)}]{Toomre_1964}
{Toomre}, A. 1964, \apj, 139, 1217, \dodoi{10.1086/147861}

\bibitem[{{Turner} {et~al.}(2017){Turner}, {Cirasuolo}, {Harrison}, {McLure}, {Dunlop}, {Swinbank}, {Johnson}, {Sobral}, {Matthee}, \& {Sharples}}]{Turner_2017}
{Turner}, O.~J., {Cirasuolo}, M., {Harrison}, C.~M., {et~al.} 2017, \mnras, 471, 1280, \dodoi{10.1093/mnras/stx1366}

\bibitem[{{{\"U}bler} {et~al.}(2019){{\"U}bler}, {Genzel}, {Wisnioski}, {F{\"o}rster Schreiber}, {Shimizu}, {Price}, {Tacconi}, {Belli}, {Wilman}, {Fossati}, {Mendel}, {Davies}, {Beifiori}, {Bender}, {Brammer}, {Burkert}, {Chan}, {Davies}, {Fabricius}, {Galametz}, {Herrera-Camus}, {Lang}, {Lutz}, {Momcheva}, {Naab}, {Nelson}, {Saglia}, {Tadaki}, {van Dokkum}, \& {Wuyts}}]{Ubler_2019}
{{\"U}bler}, H., {Genzel}, R., {Wisnioski}, E., {et~al.} 2019, \apj, 880, 48, \dodoi{10.3847/1538-4357/ab27cc}

\bibitem[{{van der Hulst} {et~al.}(1992){van der Hulst}, {Terlouw}, {Begeman}, {Zwitser}, \& {Roelfsema}}]{van_der_Hulst_1992}
{van der Hulst}, J.~M., {Terlouw}, J.~P., {Begeman}, K.~G., {Zwitser}, W., \& {Roelfsema}, P.~R. 1992, in Astronomical Society of the Pacific Conference Series, Vol.~25, Astronomical Data Analysis Software and Systems I, ed. D.~M. {Worrall}, C.~{Biemesderfer}, \& J.~{Barnes}, 131

\bibitem[{{Vieira} {et~al.}(2010){Vieira}, {Crawford}, {Switzer}, {Ade}, {Aird}, {Ashby}, {Benson}, {Bleem}, {Brodwin}, {Carlstrom}, {Chang}, {Cho}, {Crites}, {de Haan}, {Dobbs}, {Everett}, {George}, {Gladders}, {Hall}, {Halverson}, {High}, {Holder}, {Holzapfel}, {Hrubes}, {Joy}, {Keisler}, {Knox}, {Lee}, {Leitch}, {Lueker}, {Marrone}, {McIntyre}, {McMahon}, {Mehl}, {Meyer}, {Mohr}, {Montroy}, {Padin}, {Plagge}, {Pryke}, {Reichardt}, {Ruhl}, {Schaffer}, {Shaw}, {Shirokoff}, {Spieler}, {Stalder}, {Staniszewski}, {Stark}, {Vanderlinde}, {Walsh}, {Williamson}, {Yang}, {Zahn}, \& {Zenteno}}]{Vieira_2010}
{Vieira}, J.~D., {Crawford}, T.~M., {Switzer}, E.~R., {et~al.} 2010, \apj, 719, 763, \dodoi{10.1088/0004-637X/719/1/763}

\bibitem[{{Vizgan} {et~al.}(2022){Vizgan}, {Greve}, {Olsen}, {Zanella}, {Narayanan}, {Dav{\`e}}, {Magdis}, {Popping}, {Valentino}, \& {Heintz}}]{Vizgan_2022}
{Vizgan}, D., {Greve}, T.~R., {Olsen}, K.~P., {et~al.} 2022, \apj, 929, 92, \dodoi{10.3847/1538-4357/ac5cba}

\bibitem[{{Wagg} {et~al.}(2012){Wagg}, {Wiklind}, {Carilli}, {Espada}, {Peck}, {Riechers}, {Walter}, {Wootten}, {Aravena}, {Barkats}, {Cortes}, {Hills}, {Hodge}, {Impellizzeri}, {Iono}, {Leroy}, {Mart{\'\i}n}, {Rawlings}, {Maiolino}, {McMahon}, {Scott}, {Villard}, \& {Vlahakis}}]{Wagg_2012}
{Wagg}, J., {Wiklind}, T., {Carilli}, C.~L., {et~al.} 2012, \apjl, 752, L30, \dodoi{10.1088/2041-8205/752/2/L30}

\bibitem[{{Walter} {et~al.}(2022){Walter}, {Neeleman}, {Decarli}, {Venemans}, {Meyer}, {Weiss}, {Ba{\~n}ados}, {Bosman}, {Carilli}, {Fan}, {Riechers}, {Rix}, \& {Thompson}}]{Walter_2022}
{Walter}, F., {Neeleman}, M., {Decarli}, R., {et~al.} 2022, \apj, 927, 21, \dodoi{10.3847/1538-4357/ac49e8}

\bibitem[{{Wei{\ss}} {et~al.}(2009){Wei{\ss}}, {Kov{\'a}cs}, {Coppin}, {Greve}, {Walter}, {Smail}, {Dunlop}, {Knudsen}, {Alexander}, {Bertoldi}, {Brandt}, {Chapman}, {Cox}, {Dannerbauer}, {De Breuck}, {Gawiser}, {Ivison}, {Lutz}, {Menten}, {Koekemoer}, {Kreysa}, {Kurczynski}, {Rix}, {Schinnerer}, \& {van der Werf}}]{Weiss_2009}
{Wei{\ss}}, A., {Kov{\'a}cs}, A., {Coppin}, K., {et~al.} 2009, \apj, 707, 1201, \dodoi{10.1088/0004-637X/707/2/1201}

\bibitem[{{White} \& {Rees}(1978)}]{White_1978}
{White}, S.~D.~M., \& {Rees}, M.~J. 1978, \mnras, 183, 341, \dodoi{10.1093/mnras/183.3.341}

\bibitem[{{Williamson} {et~al.}(2011){Williamson}, {Benson}, {High}, {Vanderlinde}, {Ade}, {Aird}, {Andersson}, {Armstrong}, {Ashby}, {Bautz}, {Bazin}, {Bertin}, {Bleem}, {Bonamente}, {Brodwin}, {Carlstrom}, {Chang}, {Chapman}, {Clocchiatti}, {Crawford}, {Crites}, {de Haan}, {Desai}, {Dobbs}, {Dudley}, {Fazio}, {Foley}, {Forman}, {Garmire}, {George}, {Gladders}, {Gonzalez}, {Halverson}, {Holder}, {Holzapfel}, {Hoover}, {Hrubes}, {Jones}, {Joy}, {Keisler}, {Knox}, {Lee}, {Leitch}, {Lueker}, {Luong-Van}, {Marrone}, {McMahon}, {Mehl}, {Meyer}, {Mohr}, {Montroy}, {Murray}, {Padin}, {Plagge}, {Pryke}, {Reichardt}, {Rest}, {Ruel}, {Ruhl}, {Saliwanchik}, {Saro}, {Schaffer}, {Shaw}, {Shirokoff}, {Song}, {Spieler}, {Stalder}, {Stanford}, {Staniszewski}, {Stark}, {Story}, {Stubbs}, {Vieira}, {Vikhlinin}, \& {Zenteno}}]{Williamson_2011}
{Williamson}, R., {Benson}, B.~A., {High}, F.~W., {et~al.} 2011, \apj, 738, 139, \dodoi{10.1088/0004-637X/738/2/139}

\bibitem[{{Wisnioski} {et~al.}(2015){Wisnioski}, {F{\"o}rster Schreiber}, {Wuyts}, {Wuyts}, {Bandara}, {Wilman}, {Genzel}, {Bender}, {Davies}, {Fossati}, {Lang}, {Mendel}, {Beifiori}, {Brammer}, {Chan}, {Fabricius}, {Fudamoto}, {Kulkarni}, {Kurk}, {Lutz}, {Nelson}, {Momcheva}, {Rosario}, {Saglia}, {Seitz}, {Tacconi}, \& {van Dokkum}}]{Wisnioski_2015}
{Wisnioski}, E., {F{\"o}rster Schreiber}, N.~M., {Wuyts}, S., {et~al.} 2015, \apj, 799, 209, \dodoi{10.1088/0004-637X/799/2/209}

\bibitem[{{Wisnioski} {et~al.}(2019){Wisnioski}, {F{\"o}rster Schreiber}, {Fossati}, {Mendel}, {Wilman}, {Genzel}, {Bender}, {Wuyts}, {Davies}, {{\"U}bler}, {Bandara}, {Beifiori}, {Belli}, {Brammer}, {Chan}, {Davies}, {Fabricius}, {Galametz}, {Lang}, {Lutz}, {Nelson}, {Momcheva}, {Price}, {Rosario}, {Saglia}, {Seitz}, {Shimizu}, {Tacconi}, {Tadaki}, {van Dokkum}, \& {Wuyts}}]{Wisnioski_2019}
{Wisnioski}, E., {F{\"o}rster Schreiber}, N.~M., {Fossati}, M., {et~al.} 2019, \apj, 886, 124, \dodoi{10.3847/1538-4357/ab4db8}

\bibitem[{{Wootten} \& {Thompson}(2009)}]{Wootten_2009}
{Wootten}, A., \& {Thompson}, A.~R. 2009, IEEE Proceedings, 97, 1463, \dodoi{10.1109/JPROC.2009.2020572}

\bibitem[{{Zanella} {et~al.}(2018){Zanella}, {Daddi}, {Magdis}, {Diaz Santos}, {Cormier}, {Liu}, {Cibinel}, {Gobat}, {Dickinson}, {Sargent}, {Popping}, {Madden}, {Bethermin}, {Hughes}, {Valentino}, {Rujopakarn}, {Pannella}, {Bournaud}, {Walter}, {Wang}, {Elbaz}, \& {Coogan}}]{Zanella_2018}
{Zanella}, A., {Daddi}, E., {Magdis}, G., {et~al.} 2018, \mnras, 481, 1976, \dodoi{10.1093/mnras/sty2394}

\bibitem[{{Zolotov} {et~al.}(2015){Zolotov}, {Dekel}, {Mandelker}, {Tweed}, {Inoue}, {DeGraf}, {Ceverino}, {Primack}, {Barro}, \& {Faber}}]{Zolotov_2015}
{Zolotov}, A., {Dekel}, A., {Mandelker}, N., {et~al.} 2015, \mnras, 450, 2327, \dodoi{10.1093/mnras/stv740}

\end{thebibliography}
\bibliographystyle{aasjournal}


\appendix

\section{Initial values and parameter file for kinematic fitting} \label{sec:initial_values_and_parameter_file}

\begin{deluxetable}{CCCCCC}[h!]
    \tablecaption{Number of rings and initial values used in \texttt{\textsuperscript{3D}BAROLO}.} \label{tab:initial_guesses}
    \tablehead{\colhead{Source} & \colhead{NRADII} & \colhead{VROT} & \colhead{VDISP} & \colhead{PA} & \colhead{INC}}
    \startdata
    \mathrm{C1\;(A)} & 10 & 612 & 147 & 237 & 67\\ 
    \mathrm{C2\;(J)} & 8 & 212 & 65 & 110 & 49\\ 
    \mathrm{C3\;(B)} & 6 & 408 & 227 & 89 & 47\\ 
    \mathrm{C4\;(D)} & 7 & 448 & 100 & 318 & 58\\ 
    \mathrm{C5\;(F)} & 8 & 540 & 150 & 179 & 64\\ 
    \mathrm{C6\;(C)} & 6 & 218 & 96 & 316 & 46\\ 
    \mathrm{C7\;(K)} & 9 & 175 & 81 & 12 & 53\\ 
    \enddata
\end{deluxetable}
\vspace{-1.5cm}

Table \ref{tab:initial_guesses} presents the number of rings used per source, along with the initial guess for $V{_\mathrm{rot}}$, $\sigma{_\mathrm{disp}}$, $\phi$ and $i$ determined using the methods described in \cref{subsec:3Dfitting}. Tables \ref{tab:keywords_search} and \ref{tab:keywords_3dfit} show the sample parameter files used in the \texttt{SEARCH} and \texttt{3DFIT} tasks. Keywords with blank keywords are those that change with respect to the source and take values provided in Table \ref{tab:physical_params} and \ref{tab:initial_guesses}. Further information about the keyword units can be found in \citet{3db_online}.

\begin{deluxetable}{rl}[hb!]
\tablecaption{Keywords and values for the \texttt{SEARCH} task.} \label{tab:keywords_search}
\tablewidth{\textwidth}
\tablehead{\colhead{Keyword} & \colhead{Value}}
\startdata
SEARCH              & true       \\
FLAGROBUSTSTATS     & false      \\
ITERNOISE           & true       \\
SEARCHTYPE          & spatial    \\
SNRCUT              & 0          \\
THRESHOLD           & 0.00250050 \\
FLAGGROWTH          & true       \\
GROWTHCUT           & 0          \\
GROWTHTHRESHOLD     & 0.00100020 \\
MINPIX              & 1          \\
MINCHANNELS         & 5          \\
FLAGADJACENT        & true       \\
THRESHSPATIAL       & 1          \\
THRESHVELOCITY      & 1          \\
REJECTBEFOREMERGE   & true       \\
TWOSTAGEMERGING     & true       \\
\enddata
\end{deluxetable}

\begin{deluxetable}{rl}
\tablecaption{Keywords and values for the \texttt{3DFIT} task.} \label{tab:keywords_3dfit}
\tablehead{\colhead{Keyword} & \colhead{Value}}
\startdata
3DFIT      & true              \\
NRADII     &                   \\
RADSEP     & 0.0968            \\
XPOS       &                   \\
YPOS       &                   \\
VSYS       &                   \\
VROT       &                   \\
VRAD       & 0                 \\
VDISP      &                   \\
INC        &                   \\
PA         &                   \\
Z0         & 0.029             \\
FREE       & vrot vdisp pa inc \\
MASK       & search            \\
NORM       & azim              \\
TWOSTAGE   & true              \\
REGTYPE    & bezier            \\
REVERSE    & false             \\
LINEAR     & 0.42466           \\
SIDE       & B                 \\
FLAGERRORS & true              \\
ADRIFT     & false             \\
DELTAINC   & 5                 \\
DELTAPA    & 5                 \\
FTYPE      & 1                 \\
TOL        & 0.0001            \\
WFUNC      & 2                 \\
LTYPE      & 3                 \\
CDENS      & 10                \\
BIWEIGHT   & 3                 \\
STARTRAD   & 0                 \\
PLOTMASK   & false             \\
NOISERMS   & 0                 \\
NORMALCUBE & false             \\
BADOUT     & false             \\
\enddata
\end{deluxetable}

\section{Kinematic modeling results} \label{sec:model_residual_maps}

Here we show the results from the kinematic modeling using the \textsc{3Dfit} task. Each source is shown in a $2\arcsec \times 2\arcsec$ ($3\arcsec \times3\arcsec$ for C1) cut-out where
\begin{itemize}
    \item[-] row 1 shows the data, model, and residual moment-0 map. The black cross marks the kinematic center, and the black ellipse denotes the maximum radius of analysis. The elliptical $ 0.225\arcsec \times 0.166\arcsec$ beam is shown in the bottom-left corner. The red contour is at $0.2\% \mathrm{[C\textsc{ii}]_{peak}}$. 
    \item[-] row 2 shows the data, model, and residual moment-1 map. Here, the moment-1 map is the line-of-sight rotation velocity map corrected for the systemic velocity with respect to $z = 4.303$. The white circles mark the kinematic major axis of the galaxy ($\phi$). 
    \item[-] row 3 shows the data, model, and residual moment 2 map, not corrected for beam smearing.
    \item[-] row 4, columns 1 and 2 are the residual major-axis and minor-axis position-velocity maps, extracted along $\phi$ and $\phi-90\degr$ respectively. The blue contours are at $3\sigma_{\mathrm{[C\textsc{ii}],\,RMS}}$, and the grey contours are at $-3\sigma_{\mathrm{[C\textsc{ii}],\,RMS}}$. 
    \item[-] row 4, column 3 shows the inclination angle fitting. The first stage is shown in gray, while the second stage is in red. 
    \item[-] row 5 shows the position angle, rotation velocity and velocity dispersion fitting in columns 1, 2 and 3 respectively. The first stage is shown in gray, and the second stage is shown in red. 
\end{itemize}

In all images, North is up and East is to the left. 

\begin{figure*}[t]
	\epsscale{1.06} 
    \plotone{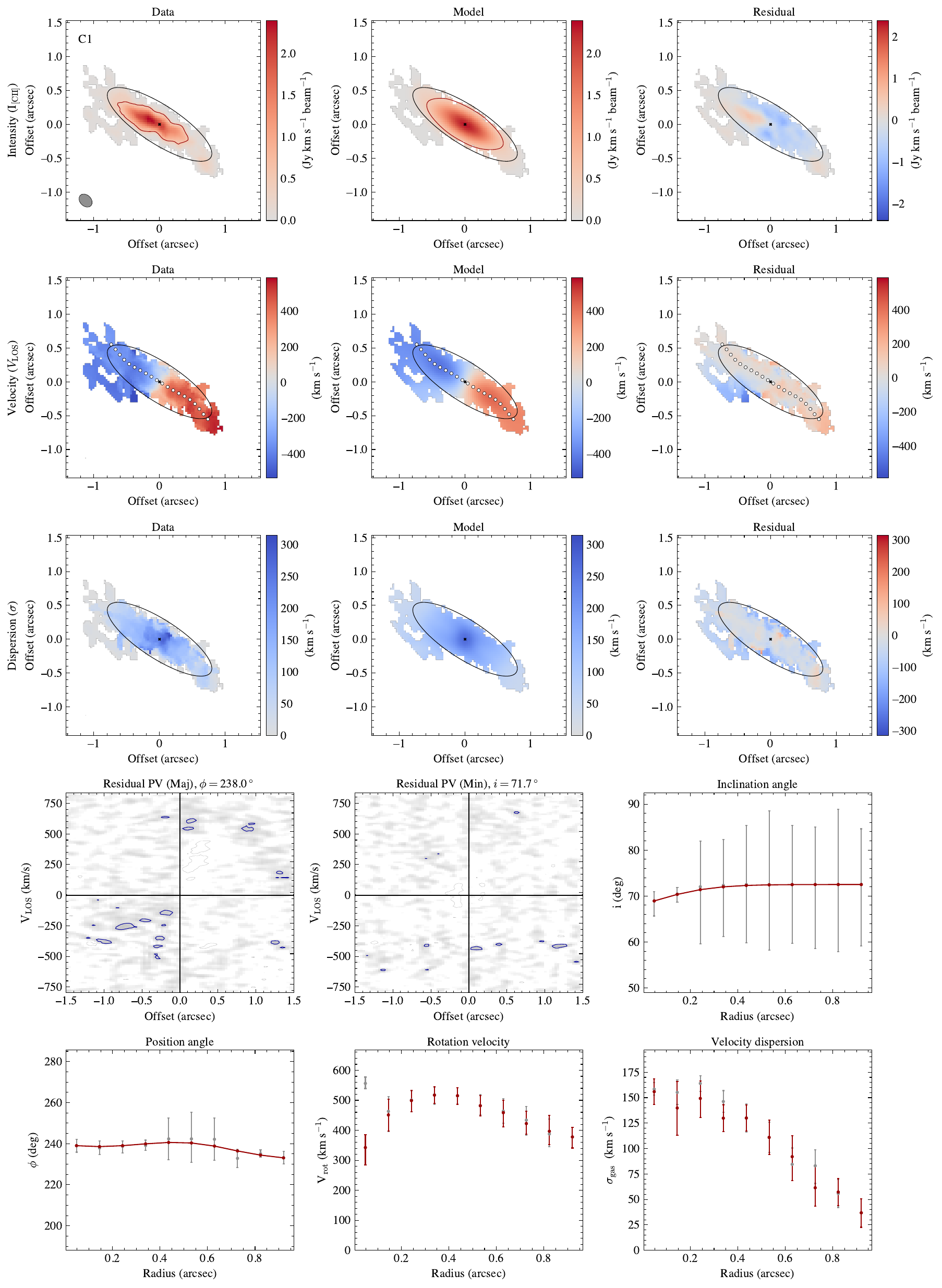}
    \caption{C1.}
	\label{fig:c1-cutouts}
\end{figure*}
\begin{figure*}[b]
	\epsscale{1.06} 
    \plotone{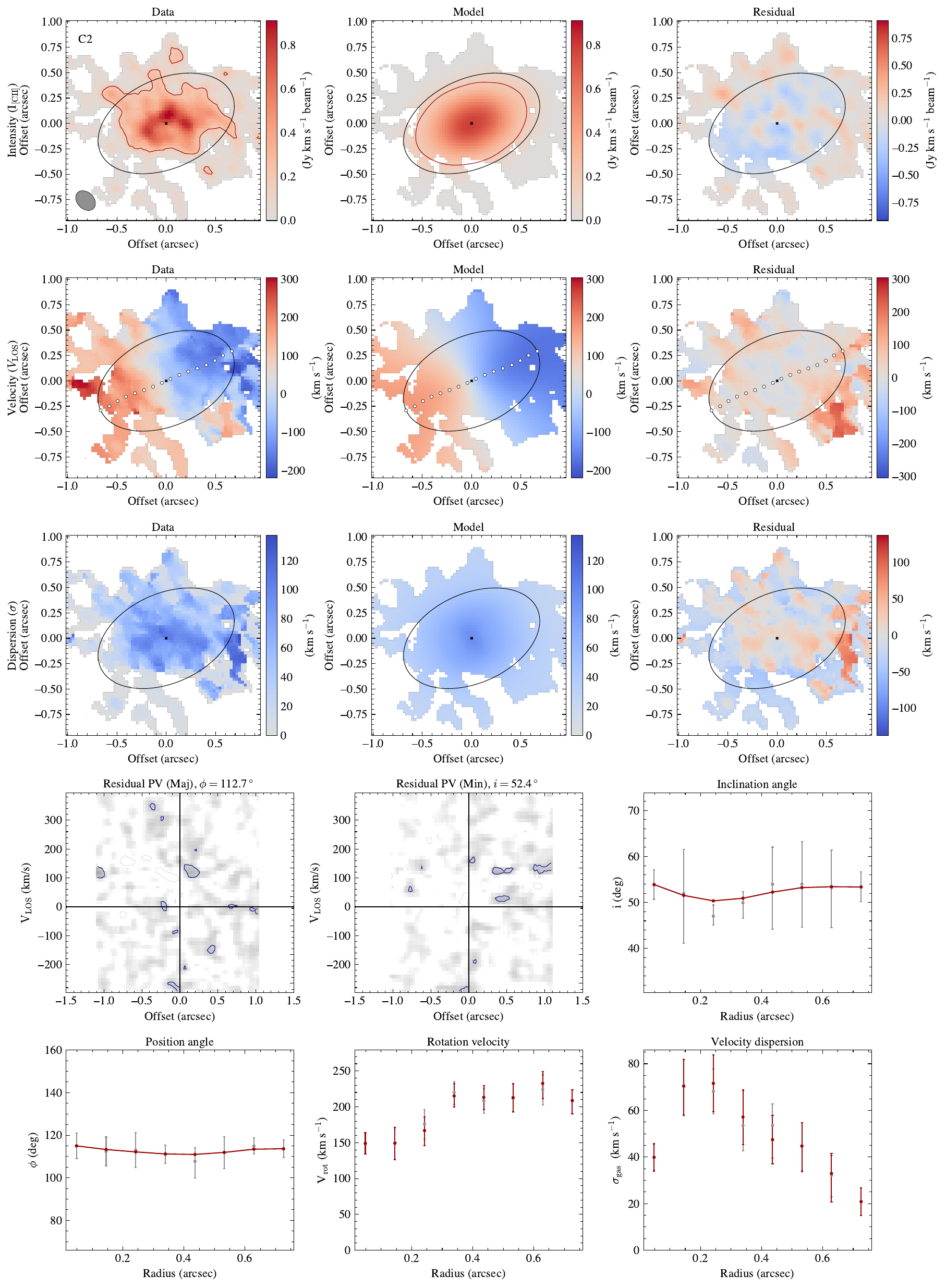}
	\caption{C2.}
	\label{fig:c2-cutouts}
\end{figure*}
\begin{figure*}%
    \epsscale{1.06} 
    \plotone{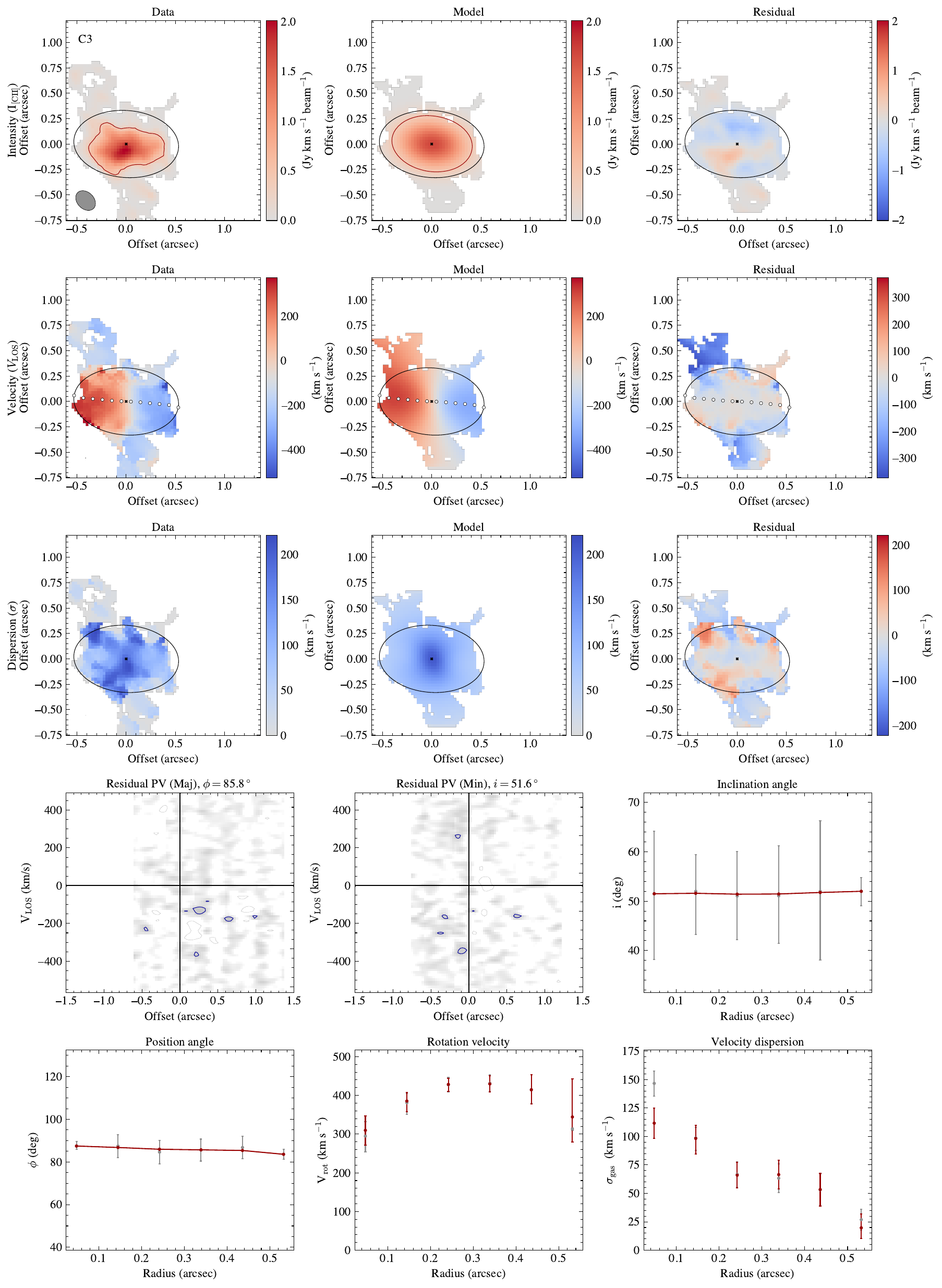}
	\caption{C3.}
	\label{fig:c3-cutouts}
\end{figure*}
\begin{figure*}
    \epsscale{1.06} 
    \plotone{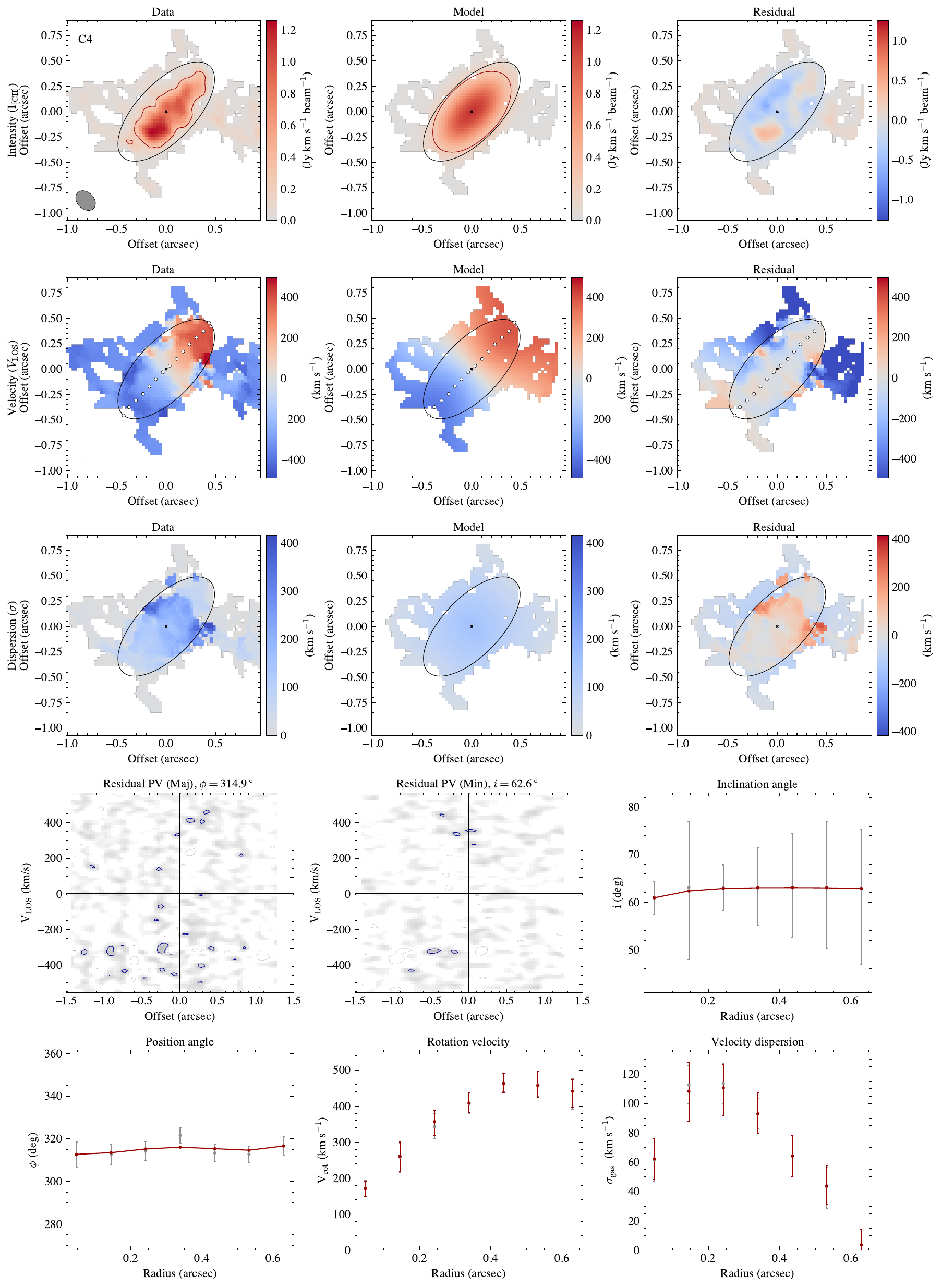}
	\caption{C4.}
	\label{fig:c4-cutouts}
\end{figure*}
\begin{figure*}%
	\epsscale{1.06} 
    \plotone{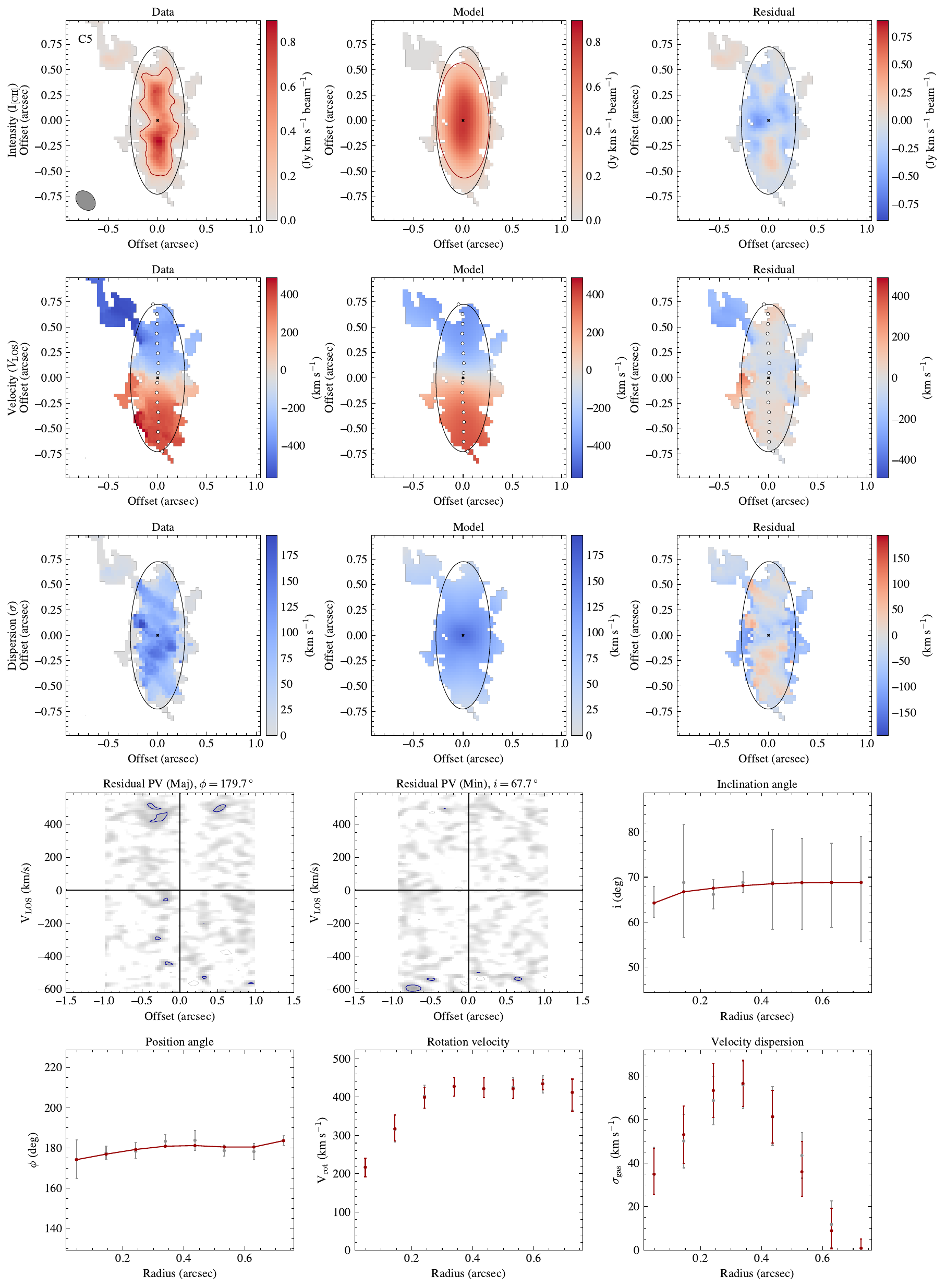}
	\caption{C5.}
	\label{fig:c5-cutouts}
\end{figure*}
\begin{figure*}
	\epsscale{1.06} 
    \plotone{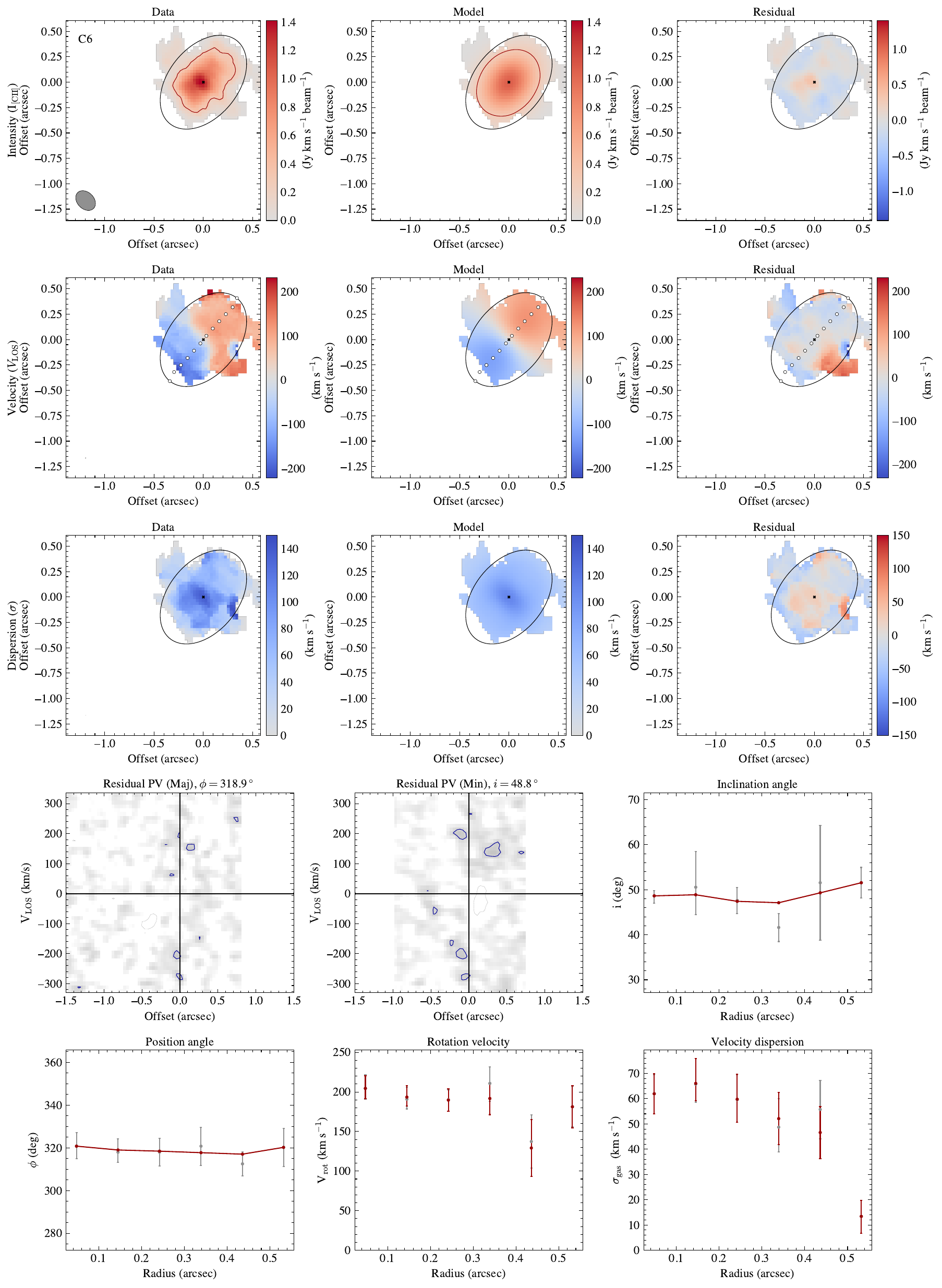}
	\caption{C6.}
	\label{fig:c6-cutouts}
\end{figure*}
\begin{figure*}
    \epsscale{1.06} 
    \plotone{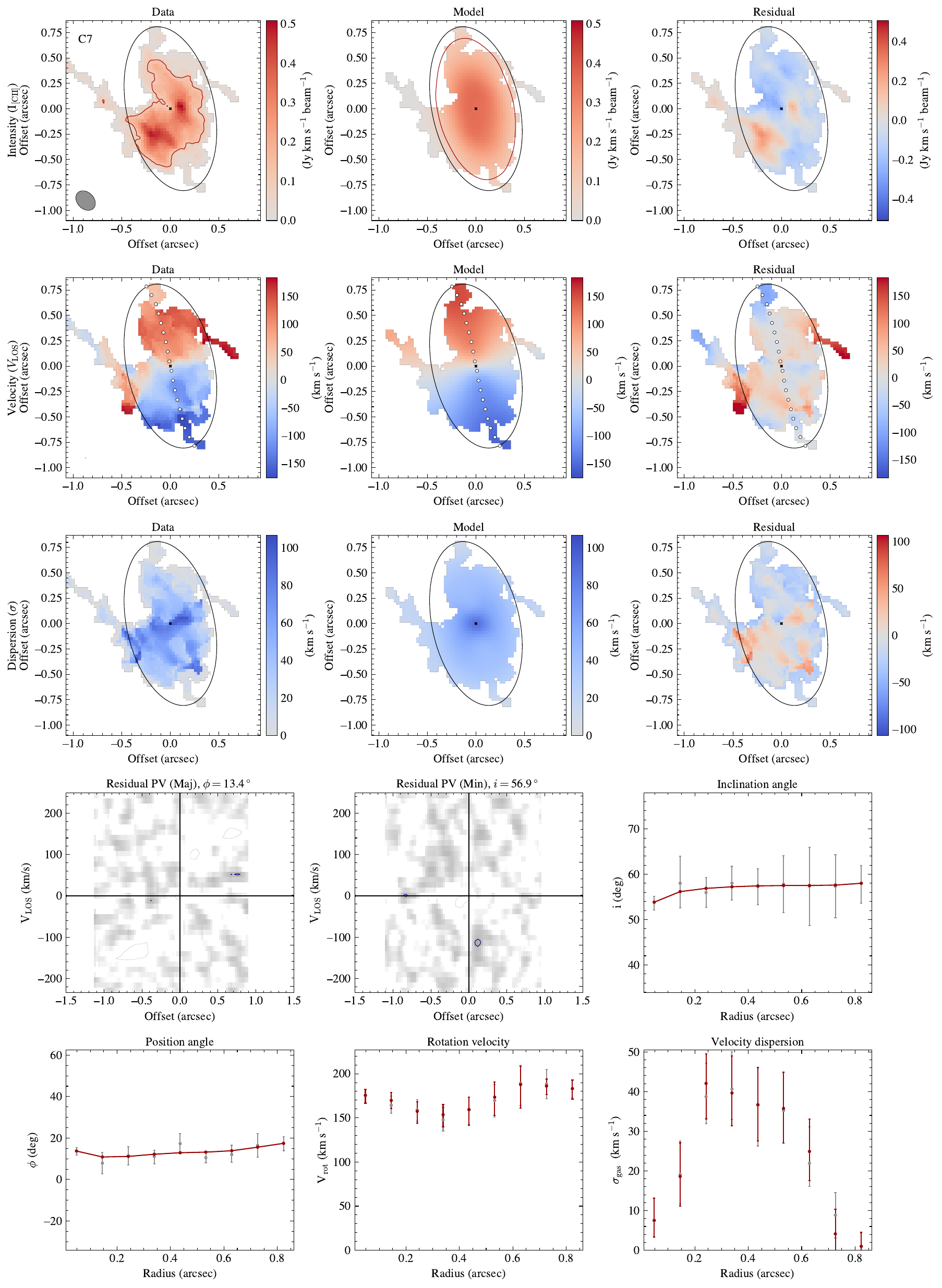}
	\caption{C7.}
	\label{fig:c7-cutouts}
\end{figure*}




\end{document}